UNIVERSITY OF ANTWERP
Faculty of Science
Department of Mathematics and Computer Science
2014-2015

# Models and Representations for Fractal Social Organizations

Arianit Pajaziti

A thesis submitted in partial fulfillment
of the requirements for the degree of
**Master of Computer Science:**
**Computernetworks and Distributed Systems**

Promoter:
Vincenzo De Florio
Co-promoter:
Chris Blondia

# Contents

# List of Figures

# List of Tables

# Chapter 1

# Introduction

Present-day technological development has reached a level where people are surrounded by digital devices and information at every moment. They have become an inseparable part of our lives. By the use of the inter-connectivity capabilities of such devices we are exposed to an extensive range of information resources and services. As a result endless communication and collaboration forms have pioneered. The newly established communication channels have become our source of knowledge, leading to new forms of socialization. This has set the ground for knowledge-based [1], cyber-physical societies [2] with enormous communication and organization opportunities. These socio-technical systems characterized by virtually instantaneous dissemination of knowledge have produced novel forms of collective intelligence and social interaction. The complexity level of such socio-technical systems or societies keeps growing in time, and as a consequence, state of the art organizational and structural approaches are in demand in order to maximize the output, while deriving efficiency.

The emergence of sought properties such as economic and social welfare for all; sustainability with respect to natural ecosystems; and especially manageability and resilience, highly depends on the way social organizations are designed. The fast growth of the human population is inducing a number of unprecedented critical situations on the natural ecosystems. In particular, natural resources are becoming scarcer, and the old ways in the organization of the human action reveal all their limitation. A more intelligent way of organization and management of resources is urgently called for. The ultimate achievement would be a society operating as a large pool of cooperating, intelligent, mobile resources. This requires learning how to engineer social-energy-aware solutions.

A typical case in point is given by traditional organizations operating in domains such as healthcare and crisis management. A common assumption characterizing those organizations is the adoption of a strict client-server model. This produces at least the following major consequences:

**Fragility.** The artificial distinction: an active and a passive side of society severely affects quality-of-emergence [3, 26]. This classification introduces penalties in the performance of the society, meaning that only a subset of the social actors is available to serve the whole set. As well known, the fast growth and the progressive aging of the human population are introducing shortages in the subset of "service providers". The problem is no more the ever increasing social costs. The management of a fragile society that is unable to serve its citizens is of the most concern. Provided the current trends the state of unmanageability is down the line.

**Absence of a referral service.** Although provided with an incomplete view of the current state of the server and its availability, it is the expected from the client to identify which server to bind to. It is the users responsibility to know, e.g., which emergency service to invoke, which civil organization to refer to, which hospital to call first, and so on. There are existing Referral services, however they mainly cover a single domain (i.e., healthcare) and very specific cases. An example observed in [4] is the seamless transfer of patient information from a primary to a secondary practitioner. Such services mostly possess an incomplete view of the available resources, because of their specialization.

**Lack of unitary responses to complex requests.** To the best of our knowledge there is no referral service that is capable of providing with composite response to complex requests such that the action, knowledge, and assets of multiple servers are combined and orchestrated in an automatic way. Even electronic referral systems in use today are mostly limited [5] and only provide predefined services in specific domains. Observed examples of such systems include

SHINE [6] and SHINE OS+ [7]. As a consequence, in the face of complex servicing requests calling for the joint action of multiple servers, the client is basically *left on its own*. Societal organizations do not provide unitary responses nor assist the client in composing and managing them. Reasons for this may be found in lack of awareness and also improper shift of service identification responsibilities from the server to the client.

The described vulnerabilities of current organizations urge the need for mutating our organizational paradigms and assumptions. An organizational approach that enables the utilization of the full potential of our societies is on demand. Consequently, the artificial distinction between an active and a passive subset should be removed or at least significantly reduced. Instead of restricting and declining the participation of roles, the organizations should promote the participation of roles, thus act as a provider and enabler. Moreover, the organization should stimulate the cooperation among the role players at all levels, from the citizens to the governing institutions. The knowledge should flow among the organization's participating entities highlighting the needs and requirements. By exchanging such information, the organization will have the capability of self-orchestrating responses, which lead to right and timely decisions.

Our society's challenge is to define one such organizational model, while preserving the fundamental aspects of the identity of the organization. A further requirement is learning how to guarantee the resilience of our evolved organizations.

A potential answer for new ways of societal organization was proposed by De Florio [8]. The proposed organizational model represents a hierarchical distributed organization of socio-technical systems, called "Fractal Social Organization" (FSO). The FSO is a nested organizational structure, built of the lower-level organizations, named Service-oriented Communities (SoC) [9]. SoCs serve as the FSOs building block.

FSO takes the responsibility as a generalized referral service for the orchestration of complex social services. In FSO, the rigid client-server scheme of traditional organizations is replaced by service orientation, while bottleneck-prone hierarchies are replaced by communities of peer-level members. Role appointment is not static and directed by the organization, but rather voluntary and context-driven. It is our conjecture that the just stated new design assumptions allow for the creation of smarter societal organizations able to match the complexity of our new complex world. In the next section we state the main goals and contributions of this thesis.

## 1.1 Contributions and goals

Our subject is oriented towards investigation of potential ways of societal organization, that allow for collective intelligent organization and management of resources. The main objective of such organizations is the exploration of the social energy from the existing societies. We conjecture that an organizational model that fulfills the above mentioned requirements is the Fractal Social Organization (FSO). Our goal is to prove and verify the effectiveness of this model by performing various simulations using the NetLogo environment[13], a tool that allows agent-based rapid prototyping. The development in NetLogo environment is based on the Logo programming language.

We begin by simulating trivial real life activities that demonstrate the main properties of the core unit of the FSO, namely the SoC. Further, more complex scenarios involving various nested SoCs are simulated. Two main simulation models have been implemented.[1] The first one simulates a virtual world, where people perform daily activities, while the normal flow is perturbed either by some natural event, which causes damage on their properties or they start having health issues, and as a result healthcare services are demanded. We compare the efficiency of the FSO approach in contrast to the Traditional social ways of organization. A detailed description of this model is provided in Chapter 4.

The second model simulates an Ambient Assisted Living (AAL) organization, where elder persons residing at their home are in need for healthcare services. By structuring additional agents based on the FSO approach we try to improve the accuracy of traditional ITC system. A detailed description is provided in Chapter 5.

The objectives of the simulations are:

- Design and implementation of FSO models that mimic social communities.
- Validation of the properties and behaviors of such organizational approaches in the healthcare domain.

[1] Link for downloading the simulation files:
`https://www.dropbox.com/s/y2vlb1ftmm5a3ns/simulation_files.rar?dl=0`

- Collection of measures and quantitative evidence of the performance of the FSO organizations based on the simulated models. In particular we are interested in service turn-around-times and service coverage.

- Assessment of resilience and trustworthiness with respect to timeliness, and availability. Goal is to verify that the strategies enacted by the FSO do not impact negatively on the design goals of existing, "traditional" organizations (e.g., the current hierarchical organization of a hospital).

The work conducted during this thesis led to the writing of two papers in close collaboration with the Promoter of the thesis. The first one with the title *"How Resilient Are Our Societies? Analyses, Models, and Preliminary Results"* has been accepted in the proceedings of the "Third World Conference on Complex Systems" (WCCS15)[2]. The second paper with the title "Tapping Into the Wells of Social Energy: An Example Based on Falls Identification" is submitted in "The 17th International Conference on Information Integration and Web-based Applications & Services" (iiWAS15)[3].

## 1.2 Structure

The thesis is organized into four main chapters. In Chapter 2 we begin by describing the main characteristics and properties of Service oriented Communities (SoC). In a separate section we describe the establishment of Mutualistic Relationships within the SoC communities. In Chapter 3 we give the main definitions of the Fractal Social Organizations and describe their structure, based on the SoCs. Further, Chapter 4 introduces the first simulation model. Here we discuss the activities/events and the implications they have on the individuals agents. The main focus is in the Health care services case study. The results of the performed experiments are provided in a separate section within this chapter. Chapter 5 describes the second simulation model, namely the Ambient Assisted Living organization and the provision of health care services to elder persons agents. After the description of the case study a detailed interpretation of the results is performed, where we discuss the extent to which our goals are met. Finally we give a conclusion of the thesis and the Future work in separate sections.

[2]Web: http://mscomplexsystems.org/wccs15/

[3]Web: http://www.iiwas.org/conferences/iiwas2015/

# Chapter 2

# Service oriented Communities

A Service oriented Community (SoC) represents a socio-technical system composed of peer entities that share services under the supervision of a coordinating agent [9]. These peer entities can be anything, including human beings, cyber-physical things etc. The participants of a SoC are named "Members". Members have distinct capabilities and goals. In [8] the characteristics of the members are named as "Viewpoints". Within a SoC the members do not have specific roles assigned, in terms of service providers or receivers. Depending on the context a member may benefit by receiving a certain service, while another one provides services. In this sense upon the occurrence of an event, members react accordingly, thus becoming active. By becoming active they play a role in the community. The service requests, or offered services are published to a service registry. The same stands for events. The service registry is handled by the coordinating member of the SoC. Based on the service registry information the coordinator is capable of estimating the capabilities and availability of community members. Thus for a given particular situation a member is assigned to play a role that enables an action. By enrolling a member for a particular action, a certain cost is estimated. Enrollments are done with a goal that might change from case to case. Sometimes a focus is given in the speed of realization, while on others safety may be the priority or cost-effectiveness. The optimization goals may be individual or social oriented. The member of the community that hosts the service registry and semantic processing can be seen as a representation of the domain of community. When an event triggers or a situation happens the coordinator is responsible for recognizing the similarities between members and binding them together. An advantage of such organization is that it brings great communication and linkage between the members, with the aim of maximizing resource usage.
As observed in [9], the above mentioned organization reminds us of Service Oriented Architecture (SOA) used in distributed systems. The participating entities or members here have the role of a service, which is able to provide or request for an action. The coordination is done depending on the situation at hand and the roles available. Figure 2.1 highlights the similarities between SoC and SOA.

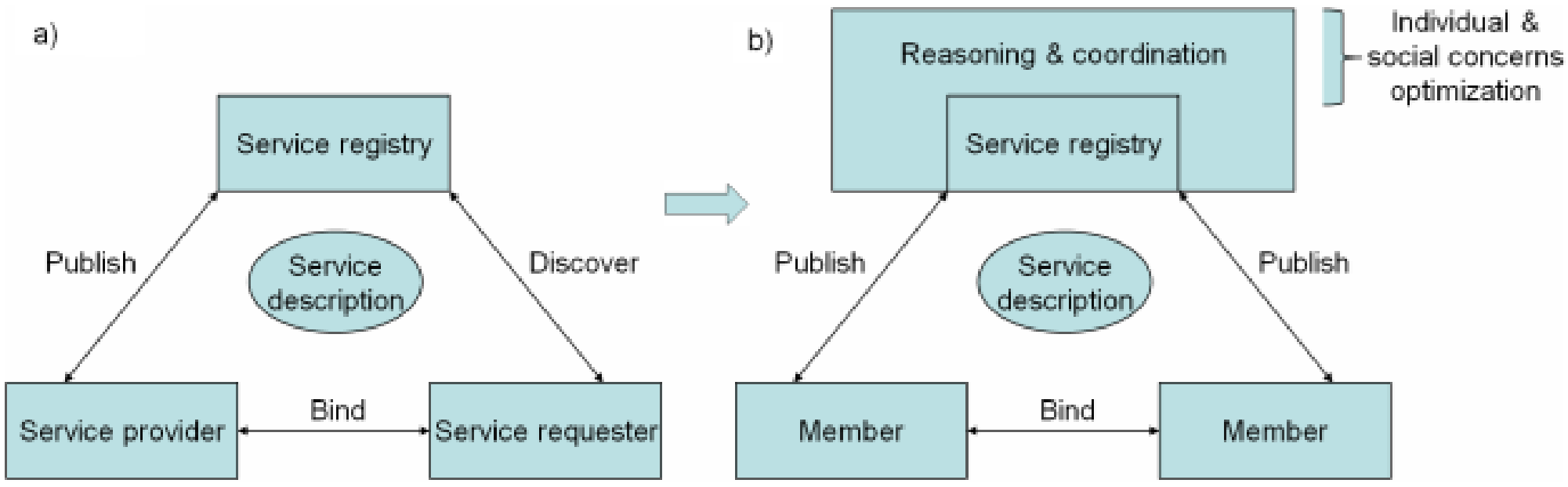

Figure 2.1: In a) the SOA architecture and b) a simple representation of SoC [9].

The interaction of different roles within a SoC creates Mutualistic relationships. This behavior is described in [12] where it was concluded that different actions performed by an entity have an effect on the domain of other entities being it positive or negative. A discussion of Mutualistic Relationships in SoC is provided in the next section. A more detailed representation of the organizational structure of a SOC is depicted in Figure 2.2.

It is observed that SoC functions as a flat society, or a cloud of social resources orchestrated by the focal point — the service coordinator. A drawback of such organization is that if the number of members that are under the management of a single coordinator becomes very large, then we would have an overload of the functions of the coordinator, which results in slow service times. Additionally, the concentration of orchestration functions on a single node results in a single point of failure issue. These issues are overcome by the Fractal Social Organizations, described in Chapter 3. FSO represent a scalable version of the SoC.

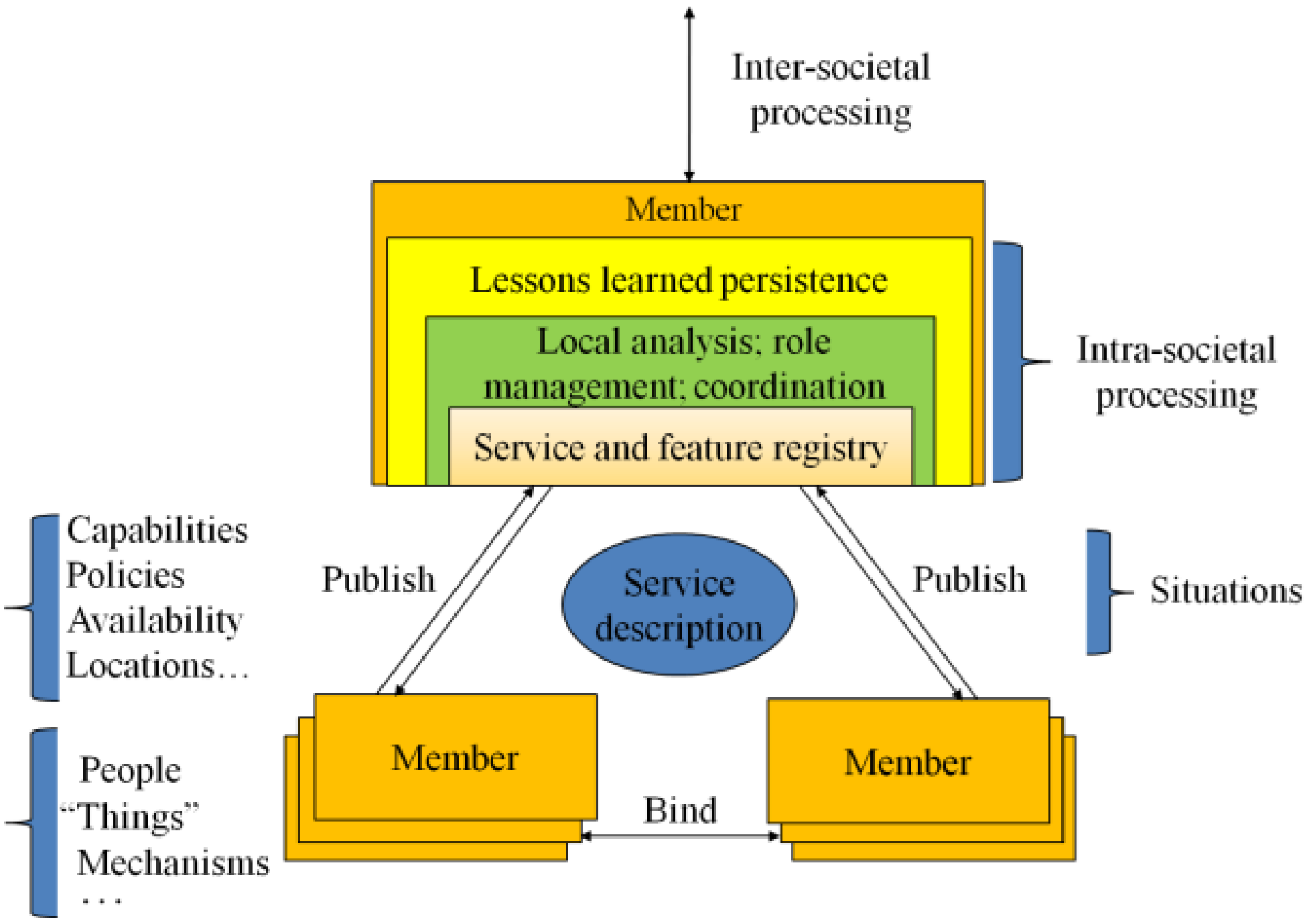

Figure 2.2: SoC representation [8].

## 2.1 Mutualistic relationships in SoC

Mutualistic relationships represent the basis of collaboration between species in nature. In a mutualistic relationship the participating entities benefit from each other. An example in nature is the one of bees and flowers. The bees fly from flower to flower to gather the nectar, necessary for nutrition of their colonies. As they land in a flower the pollen residing on the flower is attached to their hairy bodies. When the bee moves to the next flower, some of the pollen attached to their body is rubbed off their body falling on the other flower, thus pollinating the plant. The pollination is the process of pollen transfer from flower to flower allowing for plant reproducing. Like this both entities benefit from the process, the bees gather the food, while the plants get to be reproduced. This exchange of services between participating entities represents a typical case of mutualistic relationship, where each of them benefits from the process. A mutualistic relationship can involve more than one participating couple, thus creating a chain. De Florio in [12] introduces some core definitions of a mutualistic relationship model. In what follows we will describe these main definitions.

**Action function.** Given two domains or systems $X$ and $Y$, $A_X$ represents a set of actions that can be performed in $X$, while $A_Y$ corresponds to actions in $Y$. The bijective function:

$$act : A_X \to A_y \tag{2.1}$$

is a mapping of actions from domain $X$ to $Y$.

**Evaluation function.** Given a system $X$ and it's action set $A_X$ the following function:

$$eval_X : A_X \to I_X \tag{2.2}$$

maps actions, to a semantic evaluation of significance in domain $X$. The evaluation resides within at least one of the three classes: positive, meaning that the action was beneficial, neutral - action is evaluated as insignificant, and negative - action is disadvantageous.

**Mutualistic precondition.** Given two systems $X$ and $Y$, corresponding action sets $A_X$ and $A_Y$, and $I_X$, $I_Y$, the mutualistic precondition between $X$ and $Y$ fulfills the following conditions:

$$\exists a \in A_X : eval_X(a) \geq 0 \wedge eval_Y(act(a)) > 0 \tag{2.3}$$

$$\exists b \in A_Y : eval_Y(b) \geq 0 \wedge eval_X(act^{-1}(b)) > 0 \tag{2.4}$$

The first condition states that an action in $X$ is evaluated as positive and neutral, and its impact in domain $Y$ is positive, while the second condition implies the vice-versa, an action in $Y$ evaluated as positive/neutral has a beneficial impact in domain $X$.

**Mutualistic relationship.** Two domains or systems $X$ and $Y$ are in mutualistic relation, when $X$ and $Y$ trigger individual actions manifested as a form of social behavior that fulfill the mutualistic preconditions. The mutualistic relation between two domains is written as $X \mathbf{R} Y$.

**Extended Mutualistic relationship.** Finally, in [12] the notion of Extended mutualistic relationship is defined. Given two systems $X$ and $Y$ by removing two conditions from formula (2.3) and (2.4) the author describes the systems where the establishment of a mutualistic relationship, has a certain cost for the entity. The modified preconditions are modified as follows:

$$\exists a \in A_X : eval_Y(act(a)) > 0 \tag{2.5}$$

$$\exists b \in A_Y : eval_X(act^{-1}(b)) > 0 \tag{2.6}$$

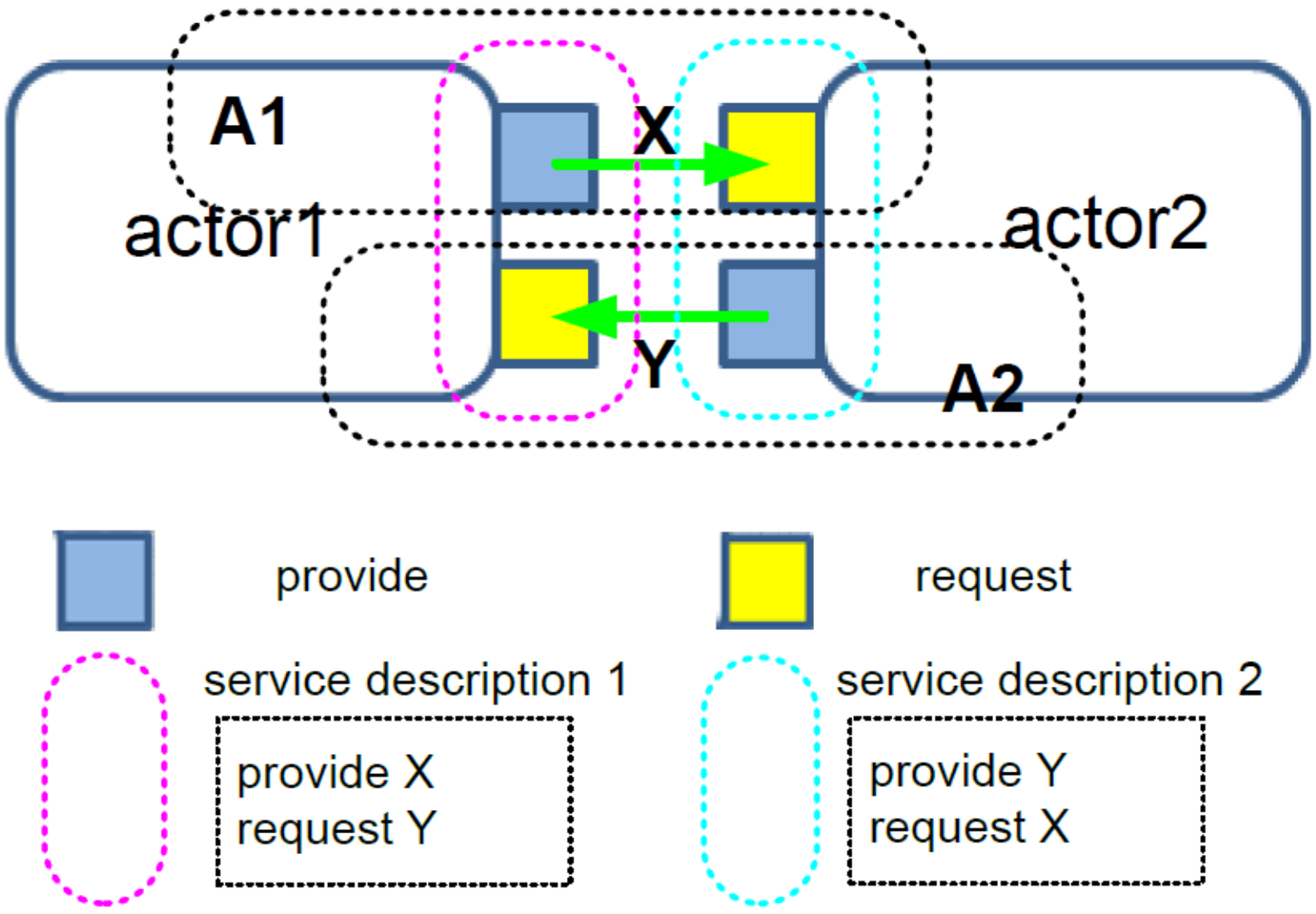

Figure 2.3: Abstract representation of a Mutualistic relationship in SoC [12].

Service oriented Communities or SoCs use semantic description of members services to allow their matching, manifested as a mutualistic or extended mutualistic relationship. In Figure 2.3 an abstract representation of a mutualistic relationship is depicted. The figure shows two actors part of two different systems. The activity *A1* triggered by *actor1* involves the provision of services that are requested by *actor2*, while activity *A2* triggered by *actor2* provisions services requested by *actor1*. If there are multiple members part of a SoC that provide the same services, and the corresponding activities fulfill each other, the activities can be merged into a so-called group activity, and the members are bound together. An example from daily life, would be the one where two individuals are willing to go for a walk with someone. Both members are providing and requesting the same services, thus they can be bound by the walking activity. This merging of two members is depicted in Figure 2.4 as a green dashed box.

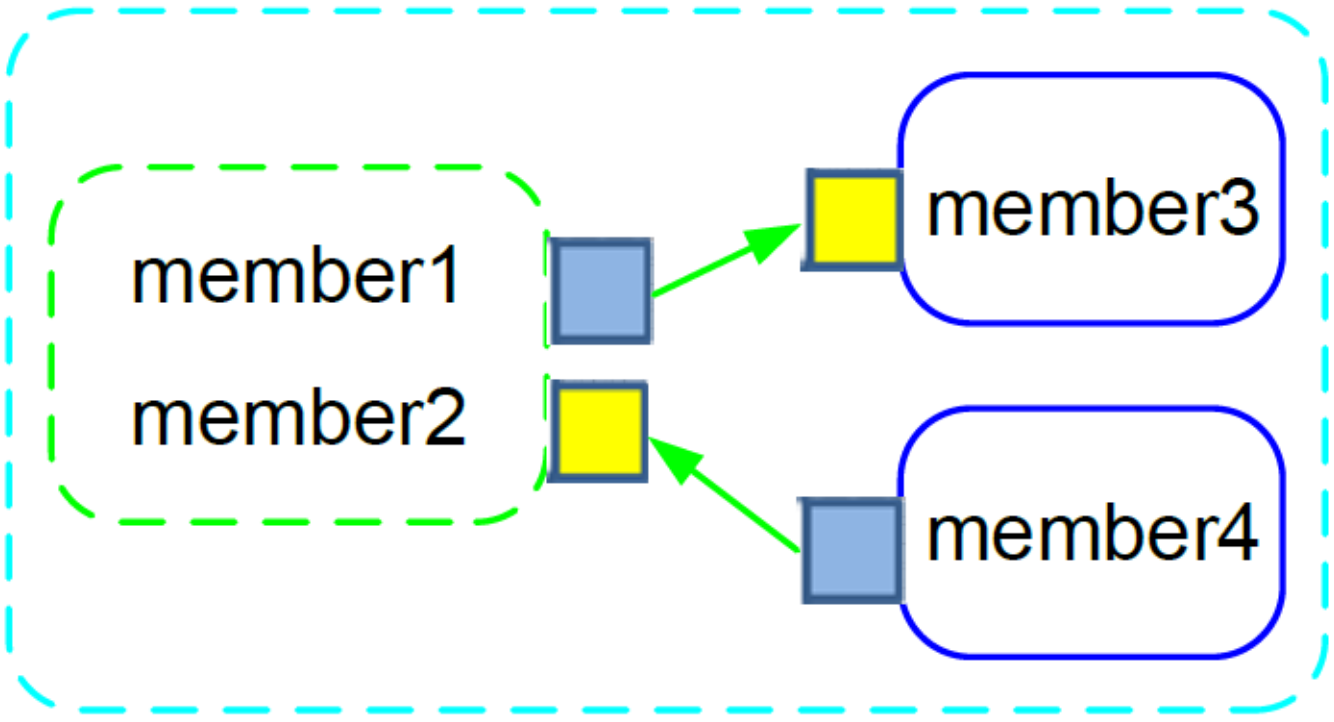

Figure 2.4: Merging of two SoC members in a group performing the same activity [12].

The newly formed member of the SoC provides walking services and is open to other members (member3 or member4 from Figure 2.4) that might request the offered service. This group of members requests for a place to go for walking, e.g. a park. A 4th member that offers such service (location) can come in help. The creation and functioning of members of a SoC as groups sets the basis for the Fractal Social Organizations. FSO are described in detail in the following chapter.

# Chapter 3

# Fractal Social Organizations

Fractal Social Organizations (FSO), introduced in [8], represent a class of socio-technical systems characterized by a distributed, bio-inspired hierarchical architecture. FSO can be described as a greater organizational structure, whose building blocks are the SoCs. The SoCs of various levels and sizes communicate with each other, thus enabling the information flow over all FSO levels. This set of rules that allows for spontaneous emergence of "social overlay networks" (SON) is called a "canon" [14], [17]. The FSO structure, can be seen as a nested SoC, whose members are other SoCs. The SoC components at the same time act as individual and social entities.

The key characteristic of FSO is the peer-to-peer approach for the collaborating nodes, meaning that based on the given situation, a node can play the role of a coordinating agent, or a simple participating member offering certain services. The communication between the various hierarchical levels is realized by the use of the so-called "exception" messages, explained more formally in what follows. These messages are triggered by the coordinators of a lower level, which make decisions whether an event should involve other entities part of the hierarchy or not. This approach allows for multiple redundant responses for a given situation. An example of FSO structural organization is observed in Figure 3.1. The SoC are placed at the layer members.

FSO allow for aggregation of services offered by humans and technological compo-nents. Given a certain event, the coordinator starts searching for roles by matching their service by semantic matching for solving the problem. Whenever the solution cannot be found an exception propagates to the next level of FSO. The procedure is repeated until a solution is found. The response results in the SON (social-overlay network)—ad hoc SoC. We now introduce the main components of an FSO.

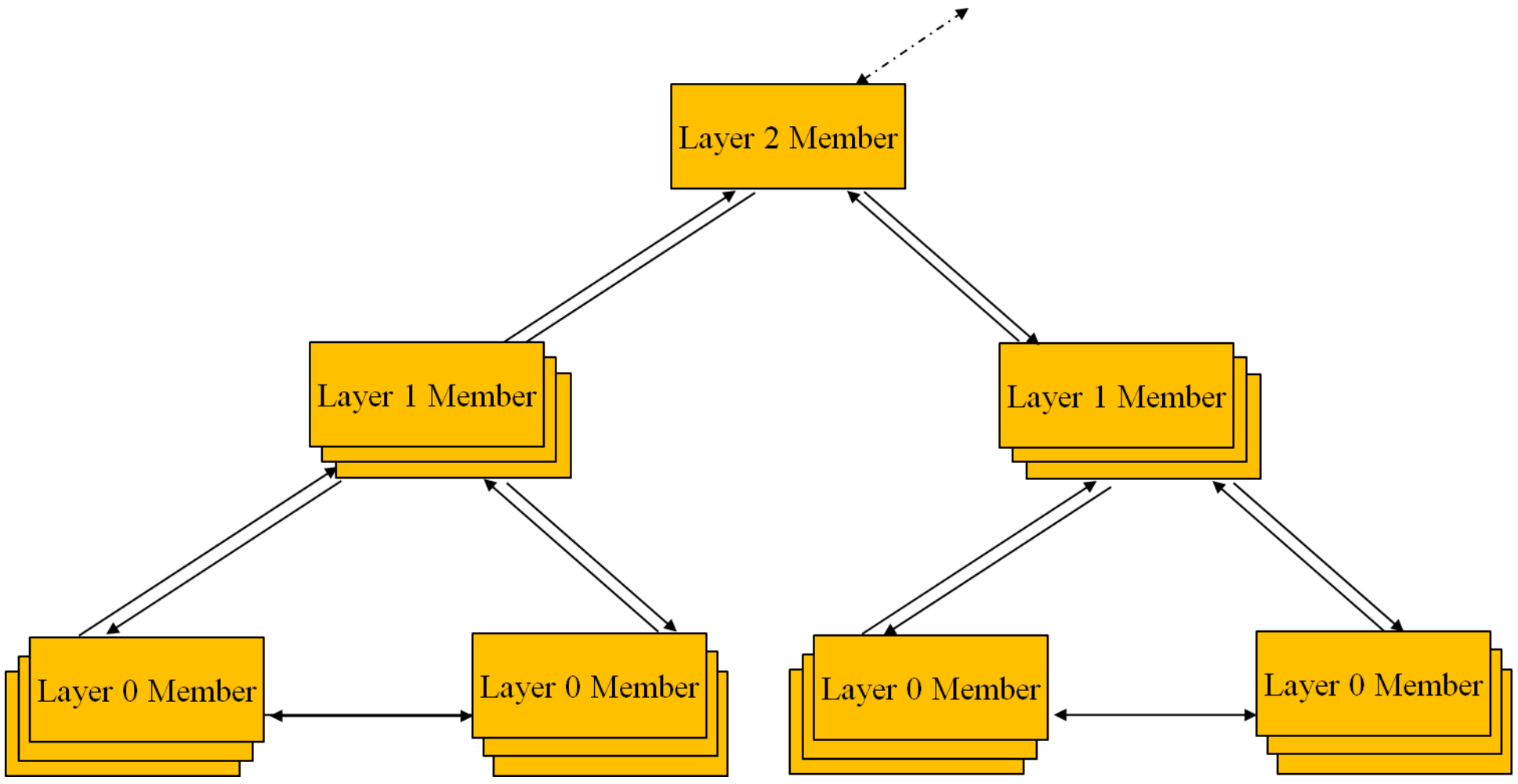

Figure 3.1: Exemplary FSO structure [8].

**Agent.** An agent is a member of the system that has an active behavior. It can be a person, a digital device, a cyber-physical system, or a collection of these entities. Agents have certain characteristics (for instance, mobility) and can be equipped with assets (e.g. objects).

**Community.** FSO may be described as a hierarchy of the Service-oriented Communities (SoC). Communities can be associated either to functional places (e.g. a flat, a building [10], a hospital) or a region characterized by the occurrence of some event, such as an accident or a fire. Seen from a higher level of the FSO a SoC is also an agent. Agents belong to communities. Based on the said definitions one might easily derive that a child community can be a participating member of another larger, parent community. This leads to the term of nested communities.

**Community representative or coordinator.** A SoC community is represented by one of its agents. This agent hosts the service registry of the SoC and is seen as the representation of the domain of the community.

If the given community is of a small size, concretely an agent together with its assets we call it an "individual SoC" (or iSoC). The community representative is at the same time a coordinator of the community it represents, and a member of the parent community.

**Role.** The Agents or members of an FSO can take a role, meaning that they perform actions that belong to a certain class of activities (e.g. doctor-, driver-, nurse-, or patient-specific activities).

**Notification.** Agents publish notifications, which can be of various natures. The notifications are received by the community representative, which reacts accordingly depending on the type of notification. A notification can be a simple message that informs about the status of an agent, a request for service, identification of a certain event etc.

**Match.** For each received notification, the community representative verifies [11] whether the new notification and those already known "enable" actions.

**Action.** Actions are performed by agents. An action represents a series of steps undertaken by the agents that initiate activities. By performing actions the agents play a certain Role within the community, thus becoming an active part of it. Every initiation of an action is followed by a notification to the community representative.

**Exception.** For a given action or event if there is a shortage of roles capable of handling it, a message is triggered from the community representative towards the higher levels of the FSO. This message is called an Exception. The exception message traverses the higher level communities until all the roles are allocated or until a flooding threshold is overcome and a failure is issued.

**Social Overlay Network.** Once we have an enabled action, the participating agents become a new temporary SoC. The lifespan of the new born SoC in limited to the duration of the activity. If the activity involves agents from distinct SoCs, the new temporary SoC is coupled of agents from different communities. The newly formed community brings together nodes from different layers of the FSO hierarchy. This temporal organization is called a Social Overlay Network (SON).

FSO is based on inter- and intra-community cooperation. This allows for establishment of mutualistic relationships between the members of within a community or distinct communities. The dynamic infrastructure of FSO allows for cooperation across the scales of the service, thus substituting the existing rigid hierarchies. The following points describe a general scenario of event handling by FSO:

- An event takes place. Participating agents of an SoC sense the change.
- The agents become active, thus implicitly play a role. The change of the status of an agent is followed by a notification published to the community representative.
- The notification reaches the SoC representative—for instance, a healthcare institution.
- The community representative handles the notification and performs local analysis for managing the response.
- An optimal response is found within the community. All roles are allocated within the current "circle", thus the response is enacted and the state is adjusted. If there is no optimal response found the following point is performed.
- Exception message is propagated requesting for roles capable of accomplishing the need. The messages traverse through the upper layers of the FSO hierarchy until the response is finally enabled or a failure is declared.
- During the process execution, new knowledge is accrued both locally and globally. The response is adjusted according to the alternatives offered at various levels of FSO.

Cases that require the above described organizational properties are quite widespread. In [26] the Katrina hurricane case that took place in New Orleans is analyzed. The occurrence of such an event led to multiple and diverse consequences. Several "social layers" were disrupted in parallel. This triggered multiple responses from the different "social layers", starting from: the Private circle (individuals, families, neighbors etc.), Private organizations, Local institutions (City Police, Fire brigades, Flood rescue etc.), State organizations (Department of emergency management), to end with the National circle. One major problem that followed the event was the coordination of these social layers. A conclusion has been derived showing that the lack of coordination between different institutional organizations of different levels lead to catastrophic consequences to the human society of that area. Moreover, the classification of the communities in informal respondents (the private cycle) and institutional respondents resulted in conflicting goals and actions. It can be easily observed that previously introduced properties of the FSO closely match the organizational and coordination requirements of the given scenario.
In the upcoming chapters we describe two simulation models that are implemented based on the given FSO concepts.

# Chapter 4

# First model

## 4.1 Overview

In this chapter, we describe one of our simulation models, its implementation and functionality. For simulations the NetLogo environment [13] is used. The simulation environment is coupled of a simulation area that functions as a “virtual world”, where the initialized agents perform their actions. It also contains a set of controls that allow for adjustment of input parameters, such as the number of agents, and various graphs that display the simulation measurements. The agents can move freely throughout the simulation area in discrete-time steps called “ticks”. A visualization of the simulation area containing the initialized agents can be seen in Figure 4.1, while the complete work-space is depicted in Figure 4.2.

In our simulation model we define SoCs of various scale and specializations, which together form a greater organizational structure, namely the FSO. Fig. 4.3 represents the hierarchical community structure used in the simulation experiments. The atomic components of the defined organization are the individual SoCs, or iSoCs. An iSoC represents a “small version” of a SoC and can be represented by a single individual (agent) and his/her personal belongings, e.g. an individual and his personal vehicle. In this case the individual is the coordinator or the representative of its iSoC, while the vehicle is a member of the given iSoC. We define agents of diverse natures including Individuals, Doctors, Firefighters, and Taxi drivers. A complete structure of the designed FSO and the participating members can be seen in Appendix A. The number of agents given in the table is not definitive, it changes depending on the simulation scenario. The emphasis is on the individuals agents and their actions or activities. By performing an activity individuals implicitly play a Role within the Community. Whenever an activity is performed by an individual, the representative of the corresponding Community records the state of the agent. In cases when there are activities that request for resources that aren't available within the community, the coordinator triggers "exception" messages to the upper level of the organization. This allows the coordinating agent to manage the underlying resources and establishment of Mutualistic relationships.

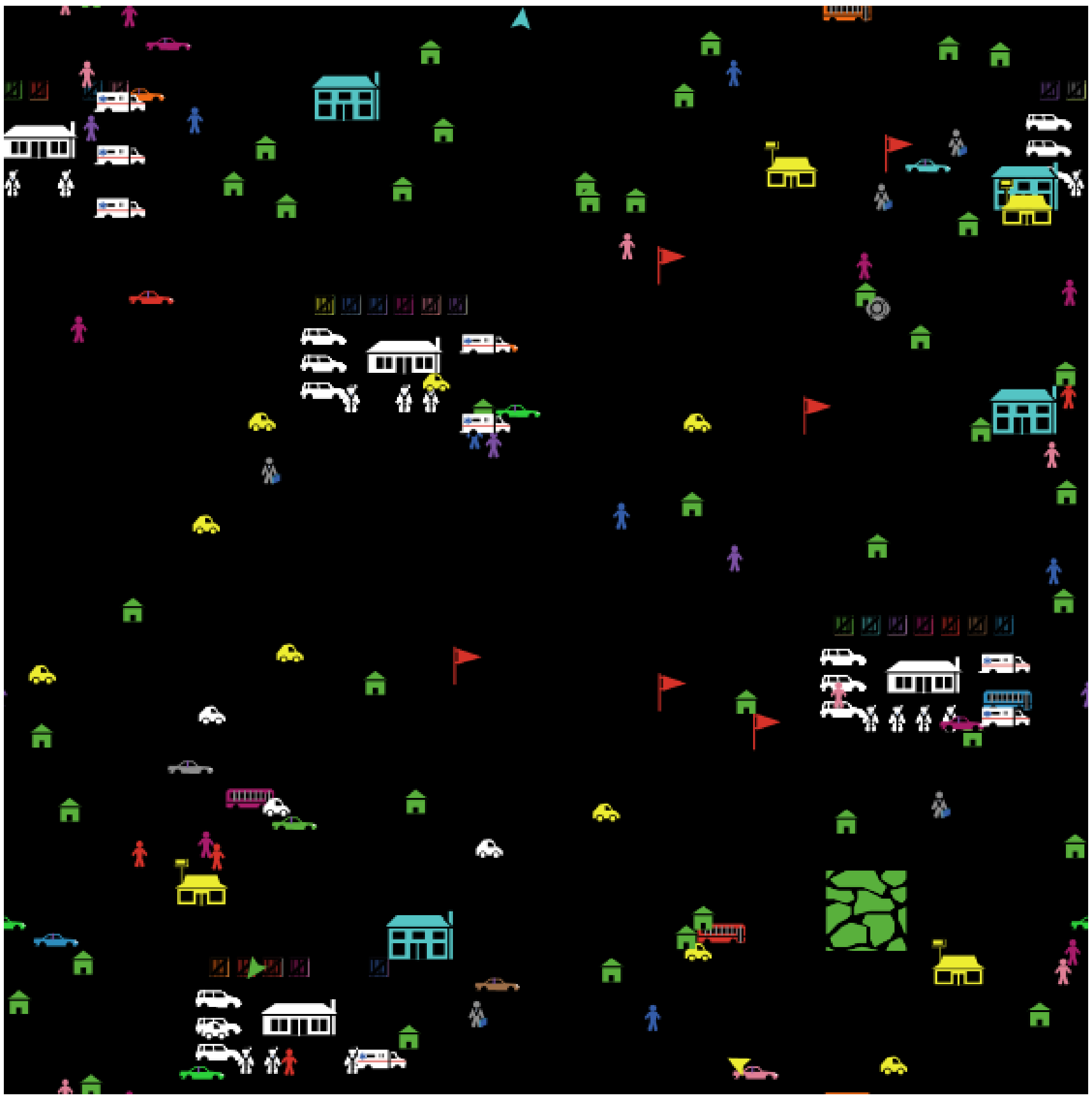

Figure 4.1: A screenshot of the simulation area, with the initialized agents.

As the simulation starts each individual starts performing an activity. The activity can be a simple process like "going to market" or a more complex one as "going to office", where the citizen has to be in office, at a certain time. For the latter, the individual has to perform time, speed and distance calculations to find out the fastest way of being there.

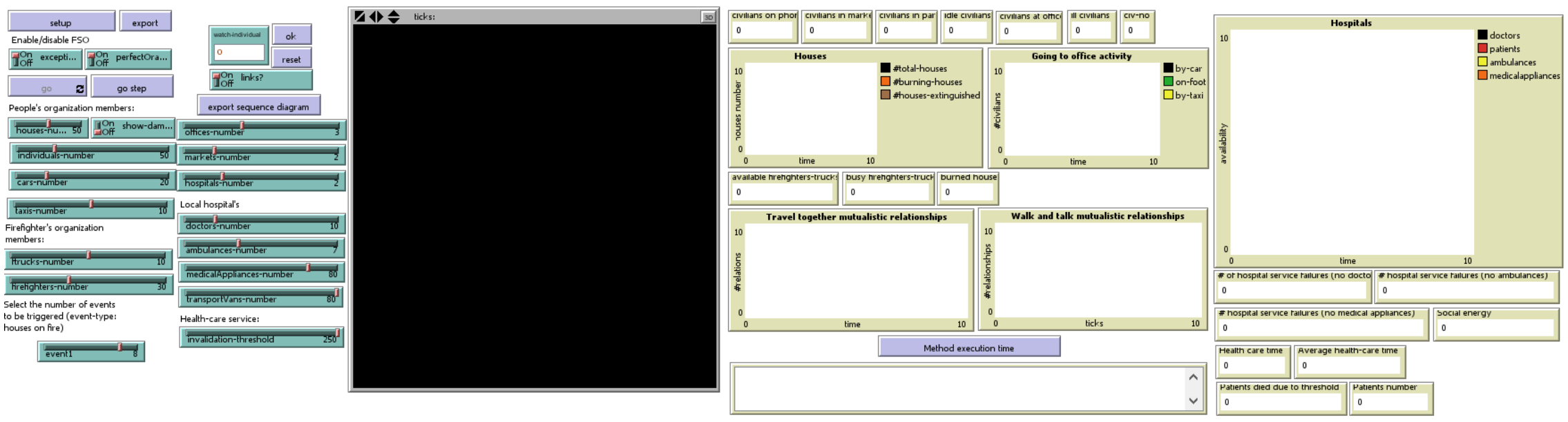

Figure 4.2: A screenshot of the work-space from the NetLogo environment used for simulations.

There are six types of activities defined. On each "tick" of the simulation if there is an idle individual, an activity is triggered for him. The activity type is chosen according to the random[1] function of NetLogo. For each activity we chose a triggering probability of 0.18, except for the "health care" activity/event for which the chosen probability is 0.09. This because it is considered that the chances for an individual to get sick or be in need of healthcare services are smaller compared to other daily activities. The activities have a certain lasting time measured in "ticks". A detailed description of each of the activities is provided in the following sections. Our focus is on the "health care" events, while the other simpler activities come in use for simulating daily life activities, and help describing the main properties of SoC and FSO. The simplest activities are those of "talk on the phone", and "go to market" which are performed solely by the individual and do not involve any collaboration between the agents. In the next section we discuss the "Walk in park" activity.

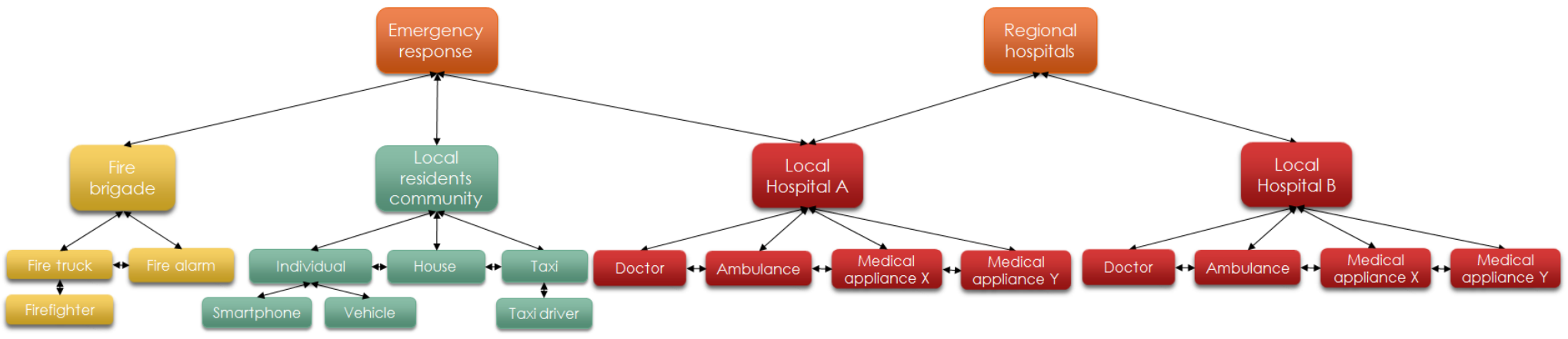

Figure 4.3: A representation of the hierarchy of Service-oriented Communities (SoC) that forms the FSO.

[1] Random numbers: http://ccl.northwestern.edu/netlogo/docs/programming.html#random

## 4.2 *"Walk in park"* activity

In Section 2.1 we gave the main properties of Mutualistic Relationships. Here we demonstrate the establishment of Mutualistic Relationships in terms of *"Walk in park"* activity.

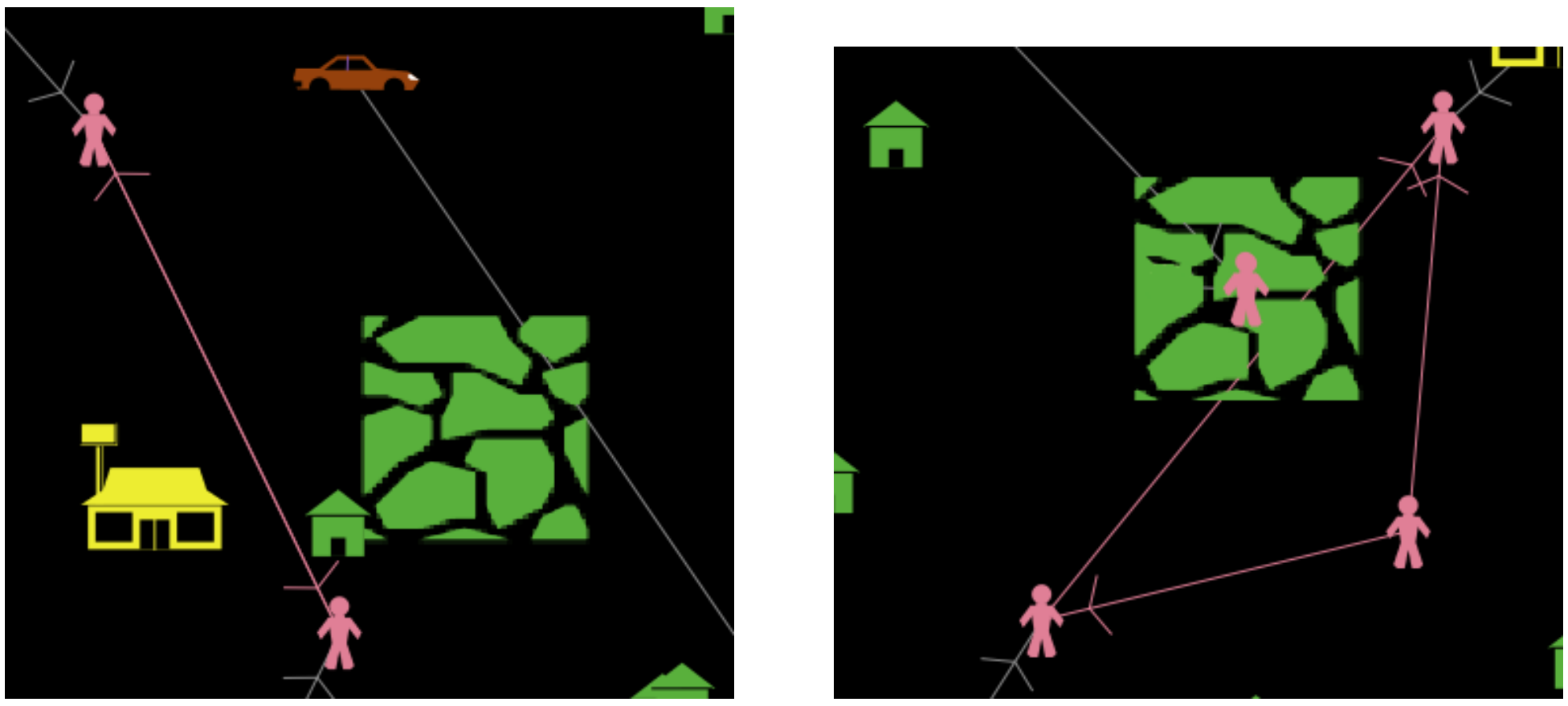

(a) Two agents (b) Multiple agents

Figure 4.4: Mutualistic relationship in the *"Walk in park"* activity.

One of the basic activities to be performed by the Individuals is going for a walk in park. The activity can be performed individually or in group. The former simulates the situation where a person decides to go walking in the park alone, while the latter simulates a group of individuals walking together in the park. In order to realize the group activity the individuals must show interest of chatting with others while walking. The group activity is defined as a composite activity, containing the chatting sub-activity. Once an individual chooses to perform such an activity, an informational message is sent to the corresponding SoC coordinator, containing information regarding the individuals preferences for the sub-activity and its current state. In cases when the individual sends a positive message, meaning that is interested to perform the activity together with someone, the coordinator starts checking for a matching activity (another individual performing the same activity, and willing to collaborate). If there is a match, a Mutualistic Relationship between both individuals is established and they proceed performing the activity together. Figure 4.4 depicts an extracted screen-shot from the simulated scenario. In Figure 4.4(a) the establishment of the Mutualistic Relationship between two individuals (agents) can be seen. The individuals performing such activity are colored in pink, same as the relation which is visualized as a two-way arrow between the two.

As described in Section 2.1 a Mutualistic Relationship can be established between two ore more agents. For the latter case, more than two individuals should be willing to perform the *"Walk in park"* activity, while chatting at the same time. An example of such scenario is depicted in Fig. 4.4b). Whenever one of the agents finishes the activity he/she proceeds with the other activities, and the links to the other agents are disconnected, thus ending the relationship.
A detailed representation of the flow of messages from the activity request sent from individuals towards the coordinator, until the establishment of Mutualistic Relationships is depicted in the sequence diagram in Figure 4.5. The green boxes represent the process where the establishment of a Mutualistic relationship is performed, while the vertical colored bold lines represent the duration of the activity.

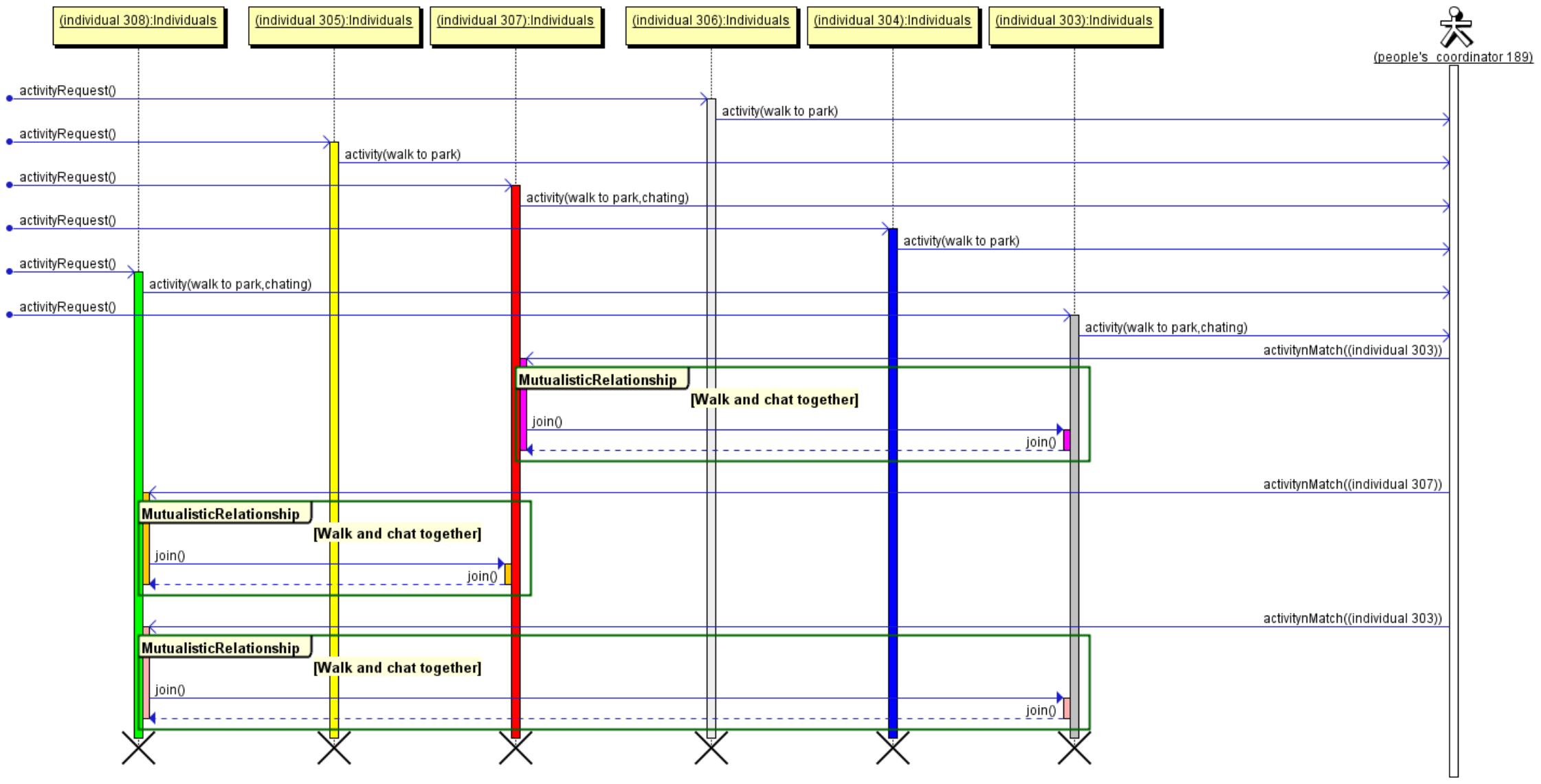

Figure 4.5: A representation of the flow of "walk in park" activities in the sequence diagram.

In the next section we describe the establishment of the Mutualistic Relationship in a scenario where not only Individuals agents are involved.

## 4.3 Car sharing activity

Another activity that can be performed by individuals is traveling to a certain location, within the two dimensional virtual world. It is assumed that the location is too far for the individual to be reached on foot, thus a car is needed for transportation. A new type of agents introduced in the simulation are the cars. An individual may or may not posses a car in his iSoC circle. In the former case he can travel to a certain destination, and also offer sharing the ride. The destination of the individual in need for transportation must be in the vicinity of the destination of the individual that is offering to share the ride in order for the activity to be performed together. Both individuals benefit from the performed activity, as one party gets the transportation, while the traveling costs are decreased.

```
(individual 193): "Exception: I need to go to x: -9.128542665351574, y: -16.667332403284007 but, I have no car!"
(individual 200): "Going to x: -11.034874855433117, y: -18.597246451651756 by car!"
(individual 196): "Exception: I need to go to x: -15.015691346723202, y: -1.194504468482279 but, I have no car!"
(individual 195): "Exception: I need to go to x: 10.031531632631555, y: 4.885535406772572 but, I have no car!"
(individual 194): "Exception: I need to go to x: 15.295837374459637, y: -15.80713565496783 but, I have no car!"
(individual 192): "Going to x: 12.614596066489753, y: 20.28507159361564 by car!"
(individual 198): "Going to x: -17.106901343618887, y: 1.8481029002817273 by car!"
(people's_coordinator 173): Mutualistic relationship: (individual 193) assigned to travel together with (individual 200), who has a car.
(people's_coordinator 173): Mutualistic relationship: (individual 196) assigned to travel together with (individual 198), who has a car.
    ......
Resource allocation time exceeded, activity: "go to location" of (individual 195) canceled!!!
Resource allocation time exceeded, activity: "go to location" of (individual 194) canceled!!!
    ......
(individual 198): "Finished the activity at location : -16.861109541760896, y: 1.5193629135608724"
(individual 196): "Finished the activity at location : -14.765276729174813, y: -1.2837576371267905"
(individual 193): "Finished the activity at location : -9.096310692953642, y: -16.686401906253575"
(individual 200): "Finished the activity at location : -10.876761185636335, y: -18.441393656406827"
```

Figure 4.6: Message exchange between the agents and SoC coordinator.

The Individuals that are to perform the car sharing activity or “go to location” action, as defined in the simulation, inform the corresponding SoC coordinator about their state. In case the coordinator finds a match, meaning that there is an individual offering to share a ride, and another one in need for transportation, the activity is performed together resulting with the establishment of a Mutualistic Relationship. We model such scenario in our simulation. The message exchange between the agents extracted from the log of the simulation is depicted in Figure 4.6.
For the individuals that aren’t in possession of a car and want to perform such an activity an invalidation threshold for the activity is defined. Once such threshold is exceeded the activity is canceled. Another type of activity can be performed afterwards. Figure 4.7 depicts the three main phases of car sharing activity.

Another representation of the flow of the car sharing activities performed by the in-

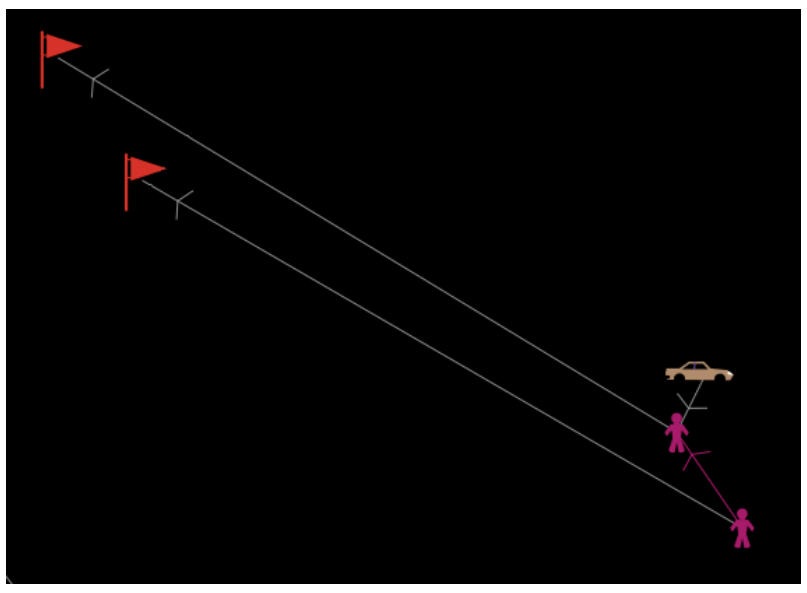

(a) The communication between the agents is established.

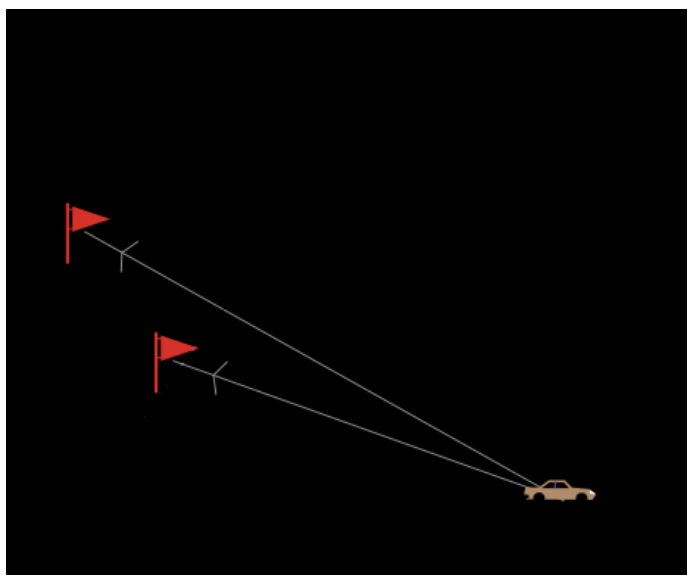

(b) The individual owning the car picks up the other Individual.

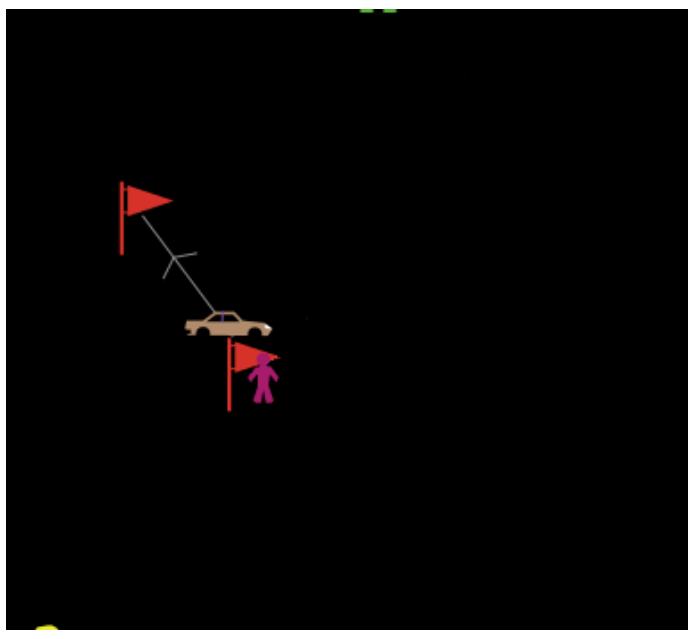

(c) The car owner drops the Individual, and continues riding towards its destination.

Figure 4.7: Mutualistic Relationship in car sharing activity.

dividuals is presented in the Sequence diagram in Figure 4.8. Same as in the previous cases the green square shows the moment of Mutualistic Relationship establishment between the two individuals, where an individual provides a certain service, which turns out to satisfy the request of another individual. The next section introduces another type of agent, which is specifically related to transportation of individuals.

## 4.4 *"Go to office"* activity

A somewhat more complex activity performed by individuals is the "go to office" activity. There are three possible ways to perform the *"go to office"* activity: on foot, by car, or by taxi. Figure 4.9 depicts an exemplary scenario extracted from the simulation where the individuals are going towards the office in one of the three above mentioned means of transportation.

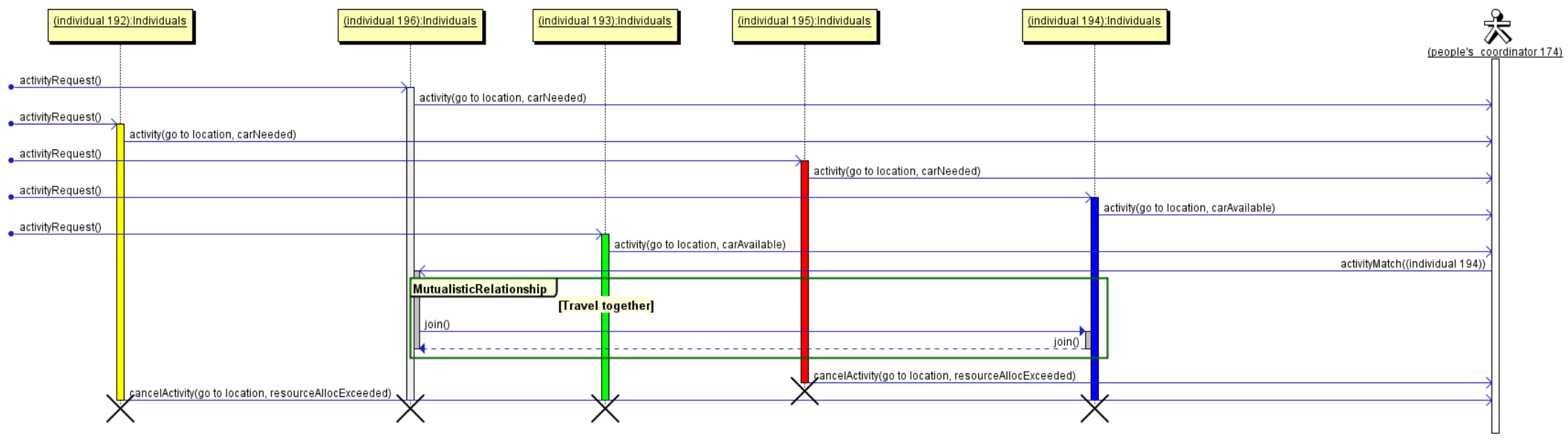

Figure 4.8: A representation of the flow of car sharing activities in the sequence diagram.

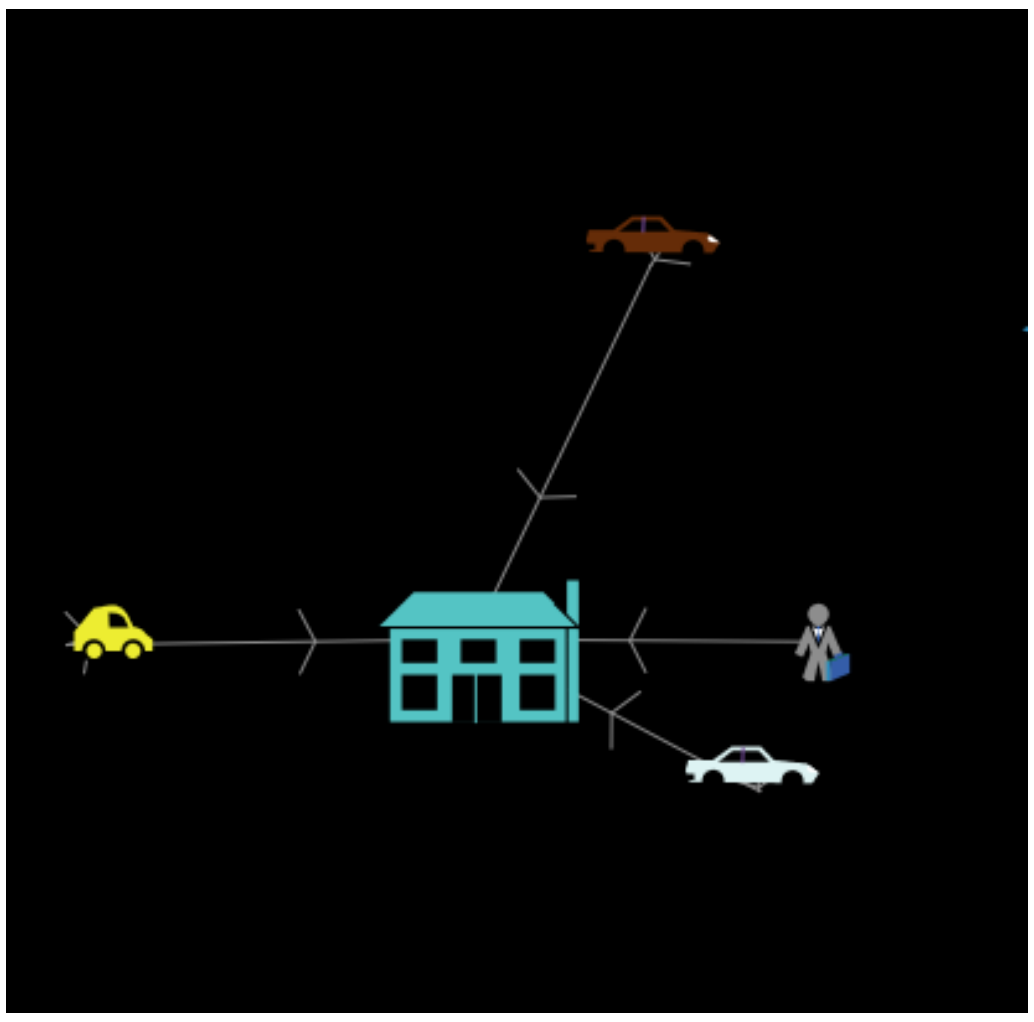

Figure 4.9: Individuals performing the *"go to office"* activity by various means of transportation.

Immediately after such an activity is triggered for an individual, a calculation of the distance from its position to the office's location is performed. It is assumed that the individuals already have knowledge about the location of their office. The activity has a defined starting time. In order to reach the office at a particular time, the individual has to choose a way of transportation that allows him to be there on time. The decision is made after a distance-time calculation is performed by the individual. In cases when there is enough time available to reach the office the individual will go on foot. This type of activity can be performed independently, thus doesn't involve any resource sharing or collaboration. In case the individual is late to walk, and possesses a car, it can be used to reach the office faster. However, some individuals lack of personal cars. In this case the exception message is raised towards the coordinator of the SoC, which tries to locate agents capable of resolving the given situation within their circle. This could be another individual, willing to share the ride (car sharing activity) or the newly defined type of agents within the SoC, the taxi-drivers together with their cars. These agents wander around the simulation area all the time, aiming to help individuals that are in need of transportation. The transport request is forwarded by the coordinator towards an available taxi. The taxi starts moving towards the individual immediately, and after picking him up they move together towards the office. Again a Mutualistic relationship is established within the Local residents circle where for a particular cost the taxi-driver helps the individual with transportation. There might be cases when no taxis are available immediately, thus the individual will be late for work, but again less time is lost. Figure 4.10 shows a sequence diagram of the *"go to office"* activities performed by the individuals.

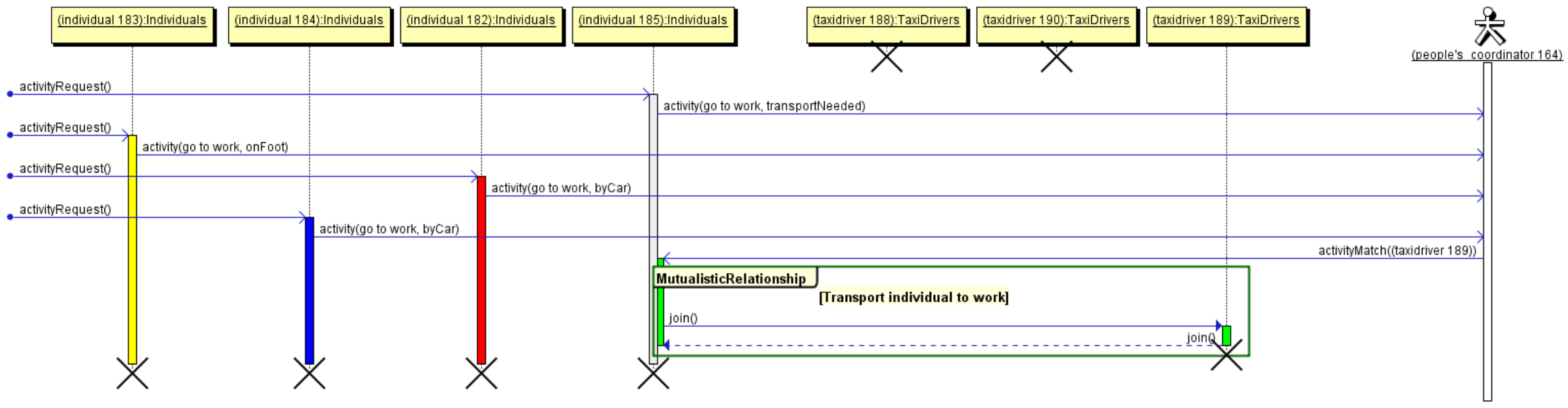

Figure 4.10: A representation of the flow of "walk in park" activities in the sequence diagram.

Running the simulation for 15000 ticks, with 4 offices, 60 Individuals, with only 15 of them owning cars, and 10 Taxi-drivers/Taxis, the results as depicted in Figure 4.11 are achieved.

It can be observed that the dominating form of transportation used by the individuals was on foot, followed by the other two forms of transportation, which have similar allocation. The scenario where the individuals are transported with the help of taxidrivers shows the social benefits of such organization. The number of times a civilian is transported by taxi represents the number of established mutualistic relationships. In the graph this is depicted by the yellow line. In the next section we describe a scenario, where the collaboration between two distinct SoCs within the FSO is involved. As we will see next, it is possible for members of distinct SoCs to establish Mutualistic Relationships.

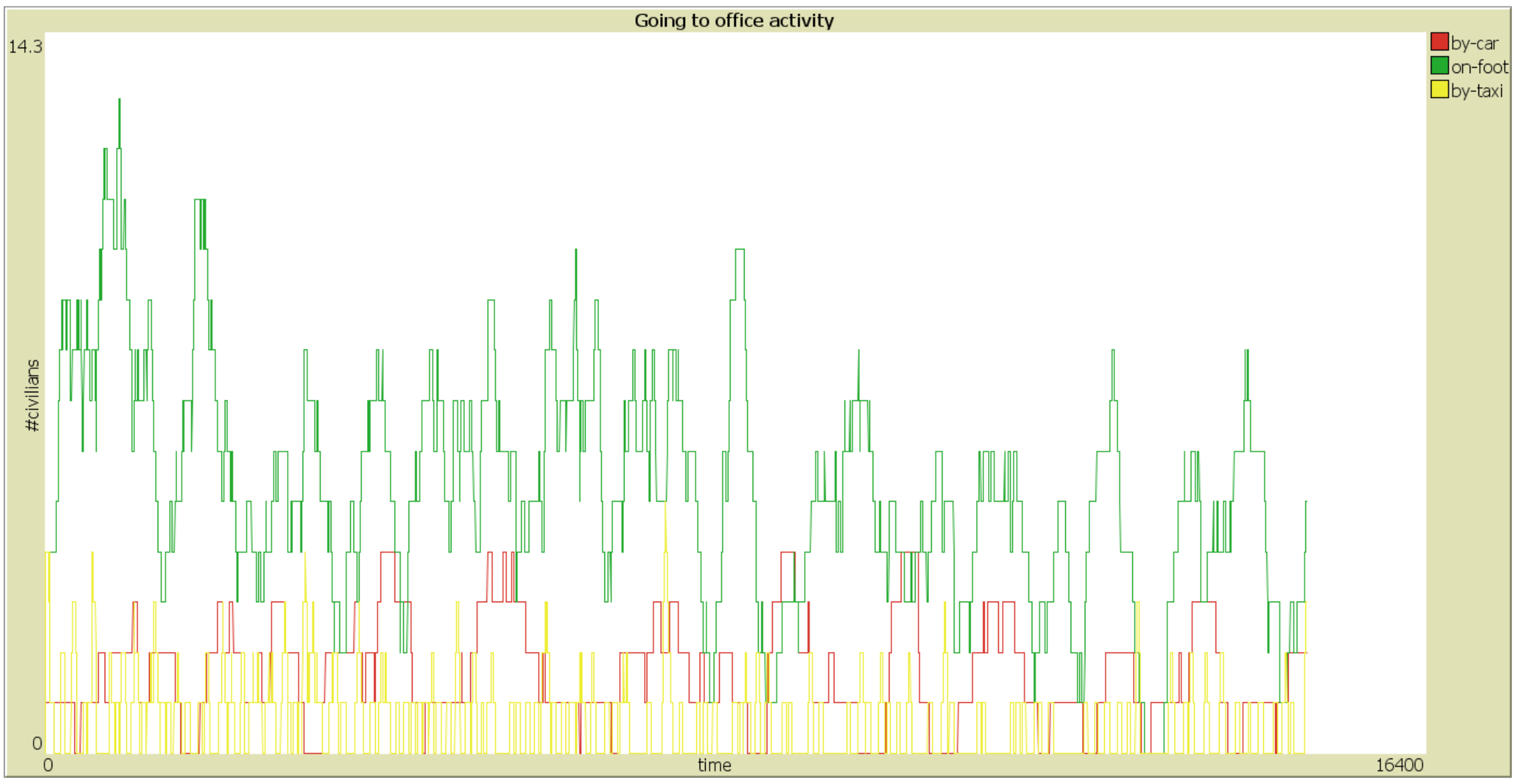

Figure 4.11: A simple graph showing the usage of various means of movement of individuals for performing the *"go to office"* activity.

## 4.5 *"Houses on fire"* event

The *"Houses on fire"* event mimics the scenario where some catastrophic natural event takes place that has an impact on the individuals properties, and as a result the houses catch on fire. In our simulator we define an event generator, that triggers "house burning" events where a number of houses start burning.
After a few cycles from the start of the simulation, some random houses will start catching on fire, thus disrupting the daily activities of the individuals. The number of events to be generated or houses to catch on fire is given initially as an input value upon initialization of simulation. The process is repeated after a certain amount of simulation cycles ("ticks") until all the houses are either completely or partially burned.
Each house has its own "health" level which starts at 100, and as it continues burning, the value decreases by a "firelevel" damage value, which initially varies within 1 to 5 range, and after each cycle it might be increased by 1 unit with probability 0.5. Once the "health" level of the house reaches 0, it is considered that the house is completely burned. The "firelevel" value is used to describe the damage that the fire is causing to the house.
Whenever there is a "house burning" event, if an individual is within the observing radius[2] of 3 of the burning house he will postpone his current activity in order to provide help. The postponed activity is continued later. In the simulation area an arrow connecting the individual to the burning house is shown and this operation will start decreasing the "firelevel" value of the house by 0.5. This mimics the "help" provided by the individual to extinguish the fire. In case the individual realizes that the fire is getting out of control ("health" level of house is below 80), and the situation cannot be handled by itself solely, an "exception" message is sent to the Local residents coordinator. Given the current situation and the lack of resources for helping for fire extinguishing the SoC coordinator forwards the "exception" towards the Level 2 Emergency response organization. Among others, this organization contains the Firefighters organization member within its circle. The complete structure of the FSO was introduced earlier in Figure 4.3.

The Firefigters SoC is composed of Fire-trucks and Firefighters. These members are also organized in iSoC fashion, where one fire-truck can hold a number of firefighters ranging from 1 to 4. The number of firefighters in a truck has an impact on the time that takes to extinguish the fire in a burning house.
The response of the firefighters organization comes immediately, by assigning a fire-truck and a certain number of firefighters to go and extinguish the fire at the given house in the "exception" message. This can be observed in the simulation area when one of the white trucks will turn red and start moving towards the house on fire (with color red). The firetrucks have greater capabilities for extinguishing fire, thus they will decrease the "firelevel" value of the burning house by a value of 2 - 5 units depending on the number of firefighters in the fire truck.
Two simulations are performed with regard to "Houses on fire" event. The number of initialized agents is as follows: 50 Houses, 50 Individuals, 10 Fire Trucks, and 35 Firefighters. The "House burning" events are generated 10 per cycle, every 100 cycles. The number of activities to be performed by individuals is six. We run

[2]Netlogo measurement unit used to measure the distance of an agent to other objects.

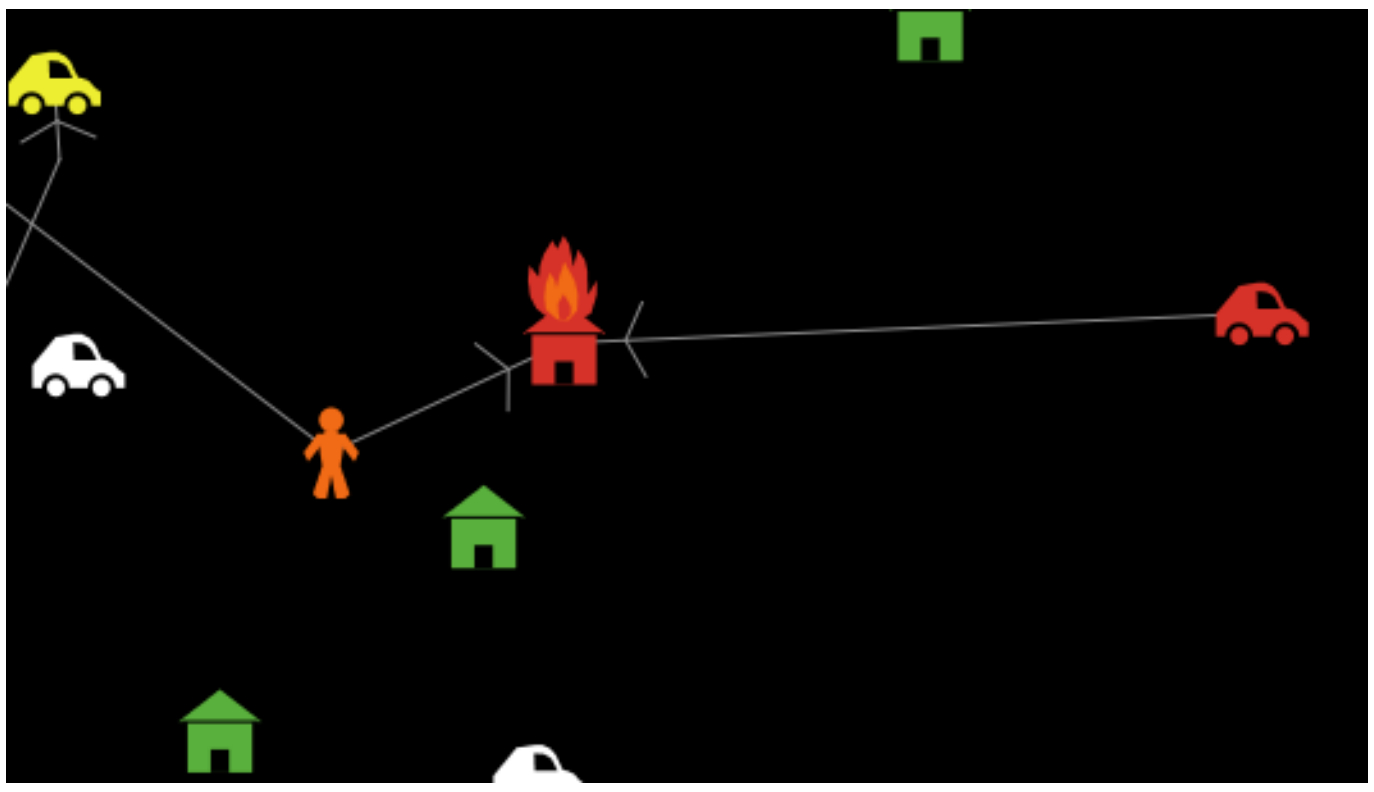

(a) A view of the individual and fire-truck helping extinguish the fire of the house.

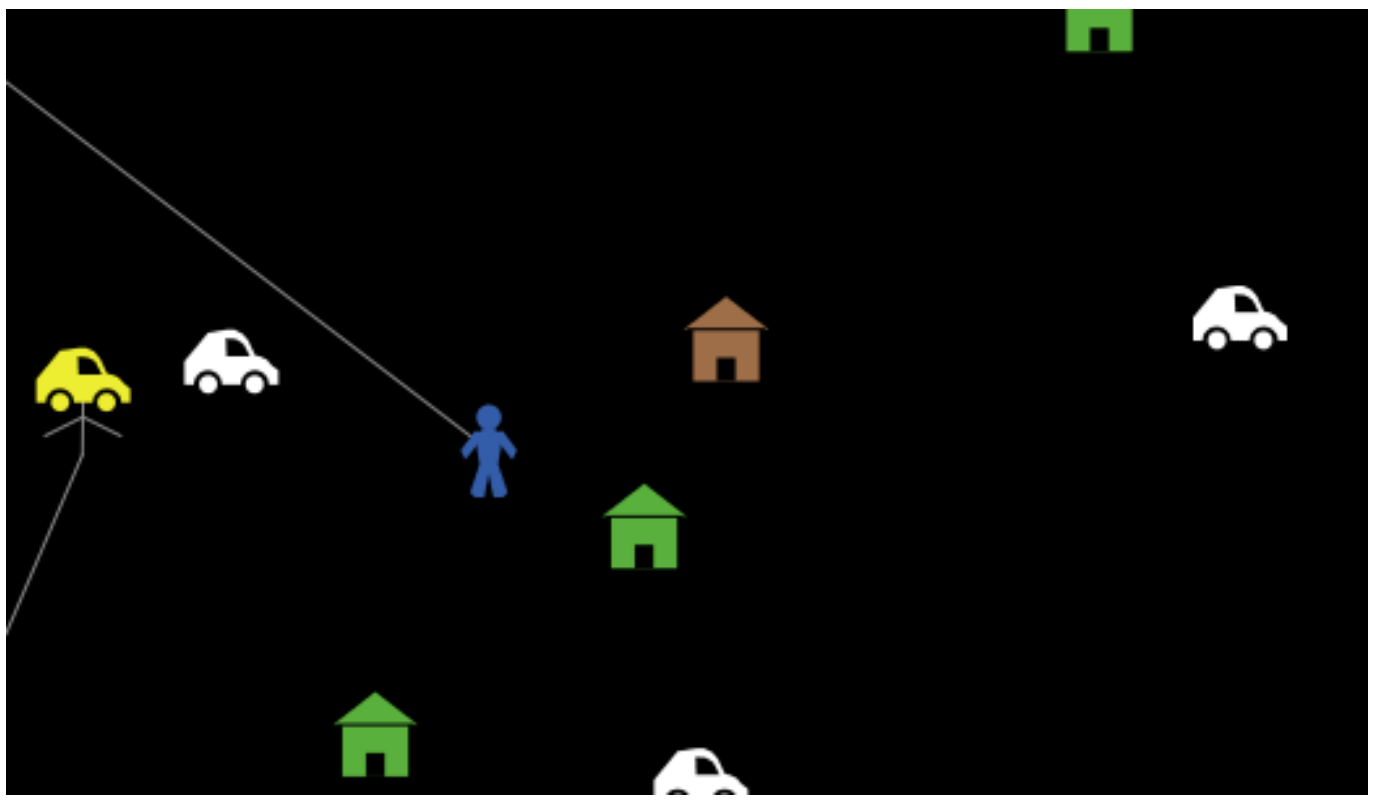

(b) A view of the scene after the intervention of the individual and the fire-truck. Fire is extinguished, the fire-truck is freed, while the individual continues his previous activity.

Figure 4.12: FSO collaboration: Agents from two distinct SoCs handling the "House burning" event.

the simulation for two scenarios: the first one there is no use of the FSO infrastructure, thus no "exception" messages are transmitted to firefighters organization and individuals cope with the triggered events solely, and in the second one the FSO infrastructure is established and the SoCs will collaborate. An extract of the simulation area showing the agents handling the *"Burning houses"* event for each scenario is shown in Figures 4.12, and 4.13. In the scenario where no FSO infrastructure is used, we see a group of three individuals are helping to extinguish the fire for the given house, thus they succeed. This isn't always the case as the number of individuals that are in the vicinity of a burning house isn't that high for regular cases, and most of the times the houses end up completely burned. Such behavior is also reflected in the simulation outcomes of the scenario explained in what follows.

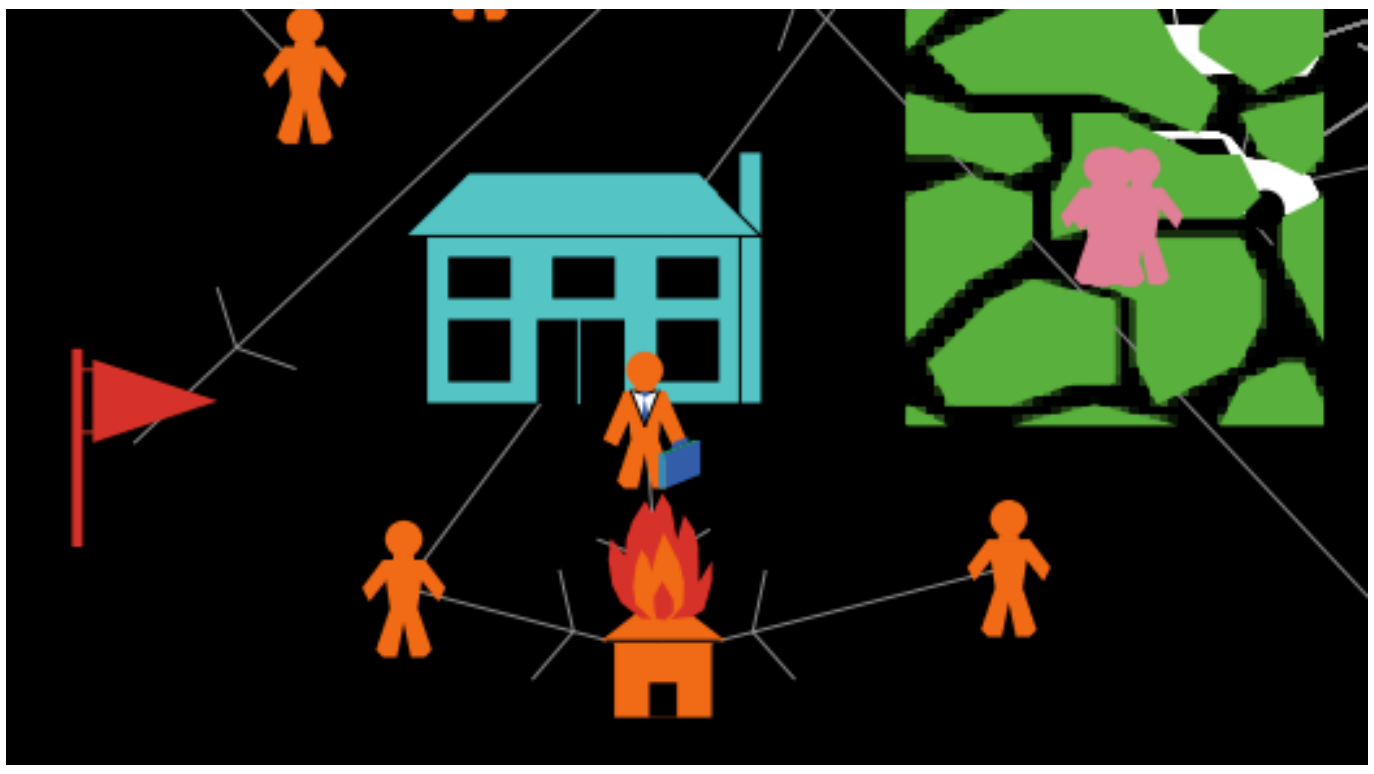

(a) A view of a group of individuals helping extinguish the fire of the house.

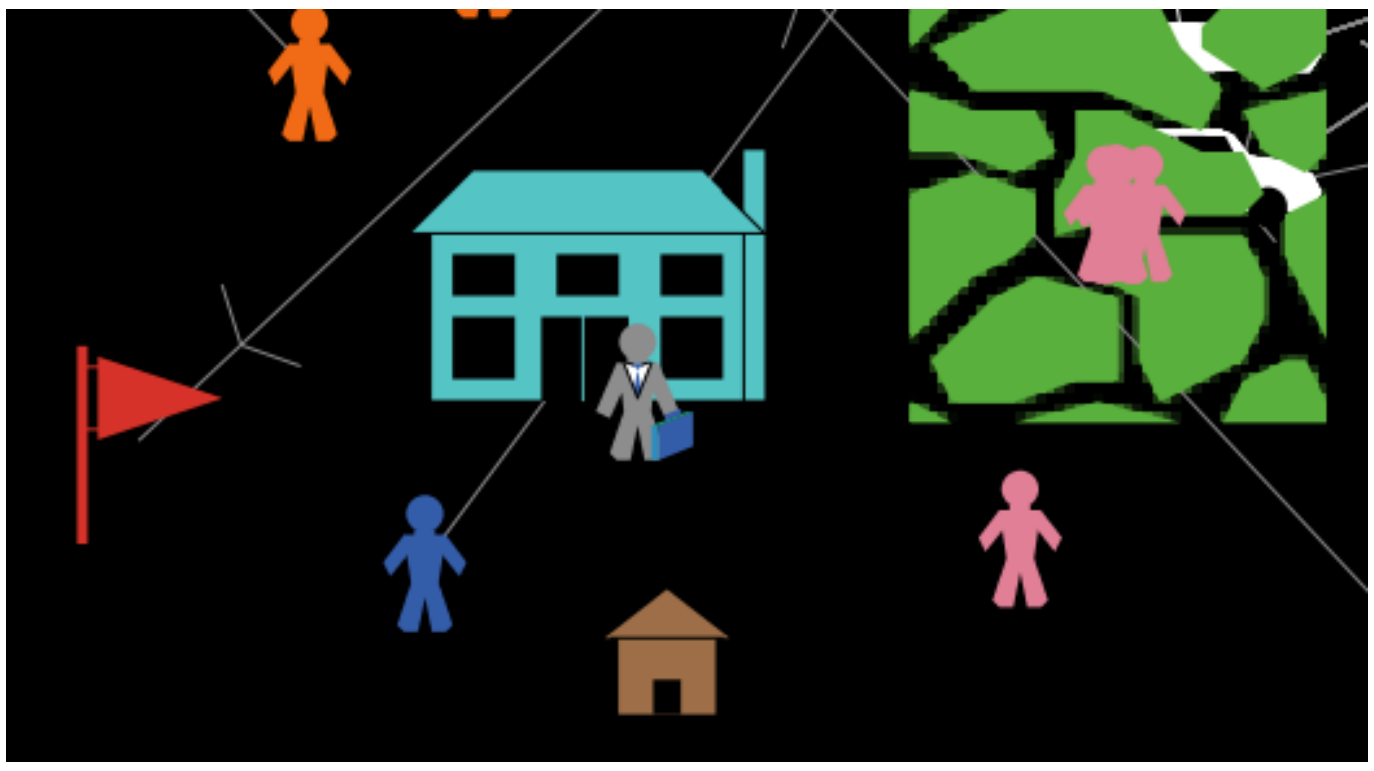

(b) A view of the scene after the intervention of the group of individuals. Fire is extinguished, while the individuals continue their previous activities.

Figure 4.13: Agents from the Local community SoC handling the "House burning" event solely.

The metric used to describe the outcome of the simulations is the total number of completely burned houses. The higher the number, the worse the result. In the case where FSO collaboration is enabled the number of completely burned houses is 29, while in the case where there is no FSO infrastructure established this number reaches 49. The results for each scenario are depicted in the graphs in Figures 4.15, and 4.14 respectively. In the first scenario we see that the "private social circle" responds to the event solely, while in the second we scenario we have a joint collaboration between the private and the "local institutions circle". By establishing a proper communication mechanism between the two level SoC communities we observe that the community benefits with less waste of resources. This reminds of the

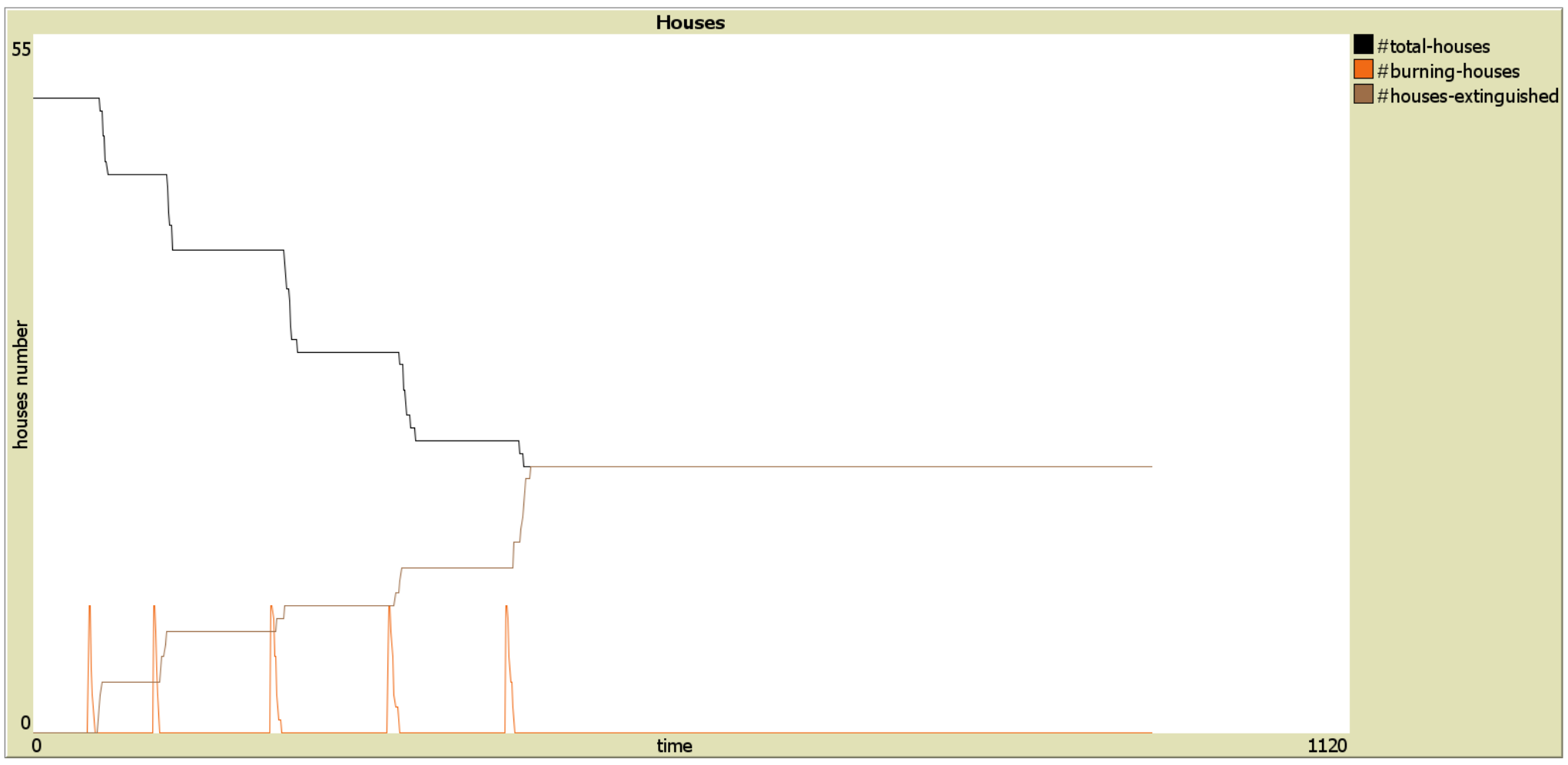

Figure 4.14: Results from the execution of *"Burning houses"* event with the use of FSO infrastructure.

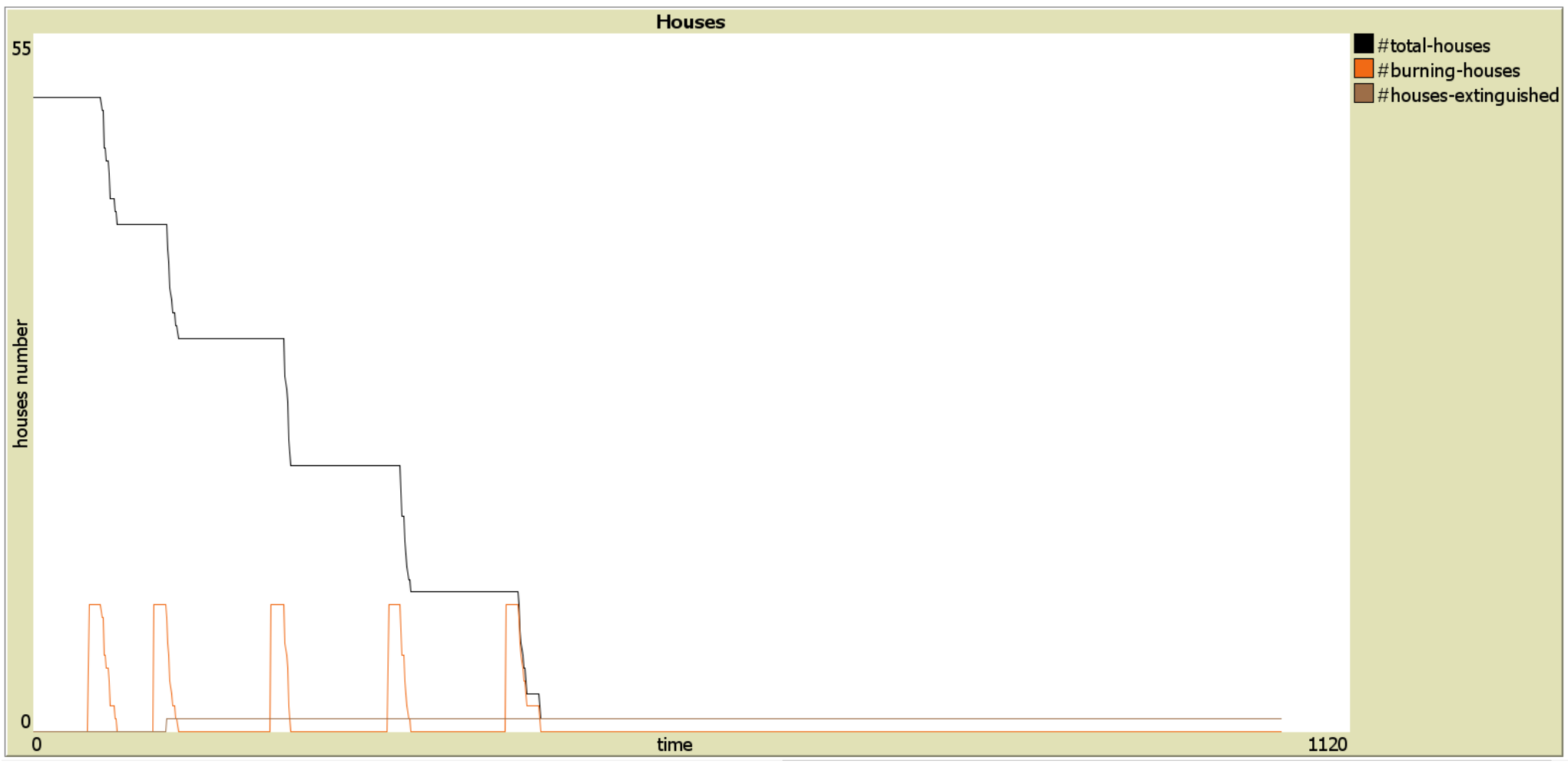

Figure 4.15: Results from the execution of *"Burning houses"* event without using the FSO infrastructure for collaboration between SoCs.

Katrina hurricane scenario described in chapter 3. It is observed that by a better organization and collaboration between the "social layers" a better management of resources is established. In the next section we discuss the Health care services case study, which represents one of our main focuses of the thesis. We show that by the full use of FSO capabilities better health care services, in terms of time management and resource allocation can be offered to individuals.

## 4.6 Case study: Health care services

A key cause of inefficiency in the interaction between users and social organizations is often the fact that said interaction is almost completely managed by the individual or by individual parts of the overall care system. The partial and outdated knowledge that the individual has about the available social resources and their state often translates in time-consuming and sequence querying or polling of services. The major added value of FSO lies in the fact that it constructs a unitary Whole that takes responsibility to assemble a coherent response to complex requests for service, thus avoiding the inefficiencies.
Based on many real-life scenarios including those described in [15],[16] the formulation of the following simulation scenarios was inspired:
Mary has just delivered her baby when she and her husband Joseph are told that the little one suffers from unexpected complications requiring specific support that is not available at the private clinic where they are now.

### 4.6.1 Traditional Case

Following the practices of traditional systems the initial step is to call the emergency service. A verbal description of the situation and the immediate requirements is provided. At the emergency service premises the case is tagged as life critical and immediate attention is taken. The operators of the emergency service might have knowledge regarding the number of hospitals in the vicinity and their specialties. However, no further information is known in terms of the hospital's resources and their availability. In search of this information, calls are made to hospitals, according to some ordered list. Through the calls, a potential hospital that might be able to handle the case is identified. An ambulance is assigned for transporting the little one and her parents to the chosen hospital. Upon reaching the hospital, the doctors discover that the situation is more complex than the way it had been described on the phone. Due to the incomplete initial knowledge, it is concluded that the needed treatment is not available at the current hospital. Calls are made to other hospitals from the list. Another one is identified as a possible candidate to supply the necessary treatment, however they lack a specific medical device crucial for the current case. A third hospital located in a further distance is reached on the phone. From the conversation it is understood that the hospital is in possession of the necessary medical device. If a complete knowledge of the resources and their availability had been available, a mutualistic collaboration between the hospitals would have fulfilled all the prerequisites for dealing with the case.

### 4.6.2 FSO Cooperation

A notification of the type "service request" is published from the emergency service operator, thus reaching the coordinator of the corresponding SoC [8, 9, 18]. Having partial knowledge about the event the SoC triggers an exception for lack of roles. The exception reaches the regional SoC which includes the originating SoC and other hospitals. At this level a more complete view of the available resources is available. The regional SoC discovers that none of the members can fulfill the request on their own. However, a combination of a set of roles from distinct SoCs (hospitals) is observed to provide with a response. The members of the identified SoCs are merged, thus becoming a SON (a temporary SoC). The new SoC elects a coordinator and enriches the request with new roles necessary to make it possible for the SON to deal with the case at hand. An ambulance is dispatched for transporting the little one and his family, while at the same time another transporting vehicle is assigned to bring the necessary medial devices to the hospital where the treatment will take place.

### 4.6.3 "Health care" event

From all the previously simulated activities the type of events or activities that we focus are the ones where Individuals are in need of health-care services. In these cases, we say that the Individual takes the role of the patient.
Throughout the simulation among the standard activities, "health care" events are triggered by Individuals, reporting that they are in need of health-care services. The severity of patient's condition in need for treatment is given as a value in the 1 - 10 range. Values 1 - 3 represent a minor illness, and it is considered that the patient can pay a visit to the local hospital by itself, whereas for values 4 - 10 an ambulance is required to transport the Individual to one of the hospitals. There are several hospitals defined, each of them in possession of a certain number of doctors, ambulances and medical appliances. A doctor can treat all three minor illnesses, and is an expert for three other diseases. In order to treat the patient the doctor uses corresponding medical appliances. In some cases the treatment of a particular condition of patient requires the use of several medical appliances simultaneously.
The request for health-care services issued by a patient must be handled within a defined threshold, otherwise it is considered that the condition degrades and results in patient's death. In our "virtual world" this case is mimicked by removing the Individual from the simulation. Based on the way the health-care request is handled we simulate three distinct scenarios, namely the Traditional Organization Case, the FSO case, and Perfect Oracle case.
In the Traditional Organization case, the individual in need for health-care services picks a random hospital and triggers the request for visit. In cases when the value of patient's condition is larger than 3 the hospital responds by sending an ambulance to transport him. Once the patient reaches the hospital a doctor is assigned to treat the patient, and based on the patient's condition the appropriate medical appliances needed for treatment are allocated. It might happen that there are no available resources within the hospital to treat the patient, e.g. no available doctors that are experts on treating a particular condition, or there are no medical appliances. In these cases a "polling" in the other hospitals is performed. The patient's request is forwarded to the next hospital, without an exact knowledge about the hospital's resource availability. The procedure is repeated until the patient reaches a hospital that is able to offer the appropriate treatment, or the request exceeds the predefined threshold, and the patient dies.

The FSO approach treats the health-care service events in a different manner. First, the Individual in need of health-care service raises an exception from its iSoC to the representative of the corresponding SoC (Local residents community). The representative or the coordinating agent handles the exception and checks for available resources able to cope with the given event within its circle. In case of a negative response the exception is forwarded to the upper level representative, namely the "Emergency response" agent. Located at a higher hierarchical level the "Emergency response" has broader knowledge regarding the agents that might be able to deal with the situation at hand. After locating "Local hospital A" (the main local hospital) as a potential helping agent, the notification is directed towards it. The "Local Hospital A" is well informed about its underlying agents, and resources. In the

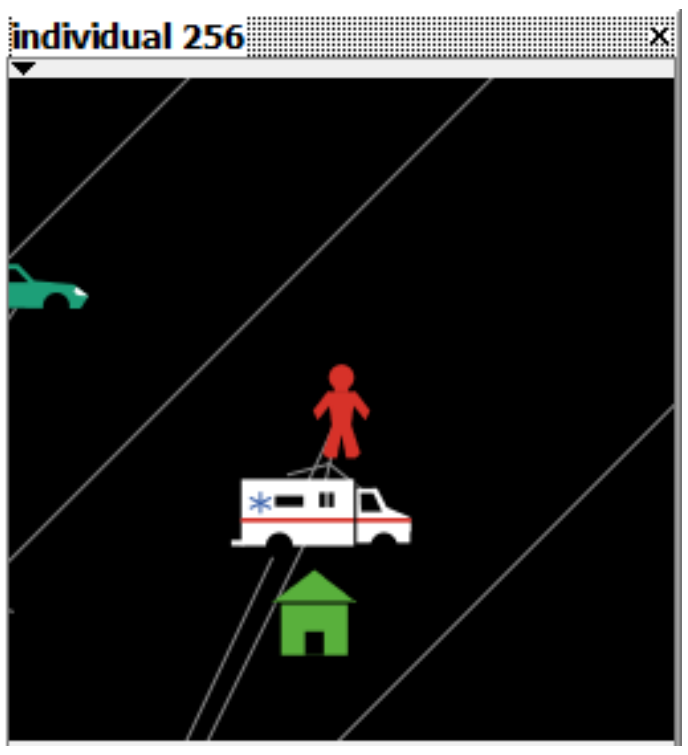

(a) Patient waiting for ambulance.

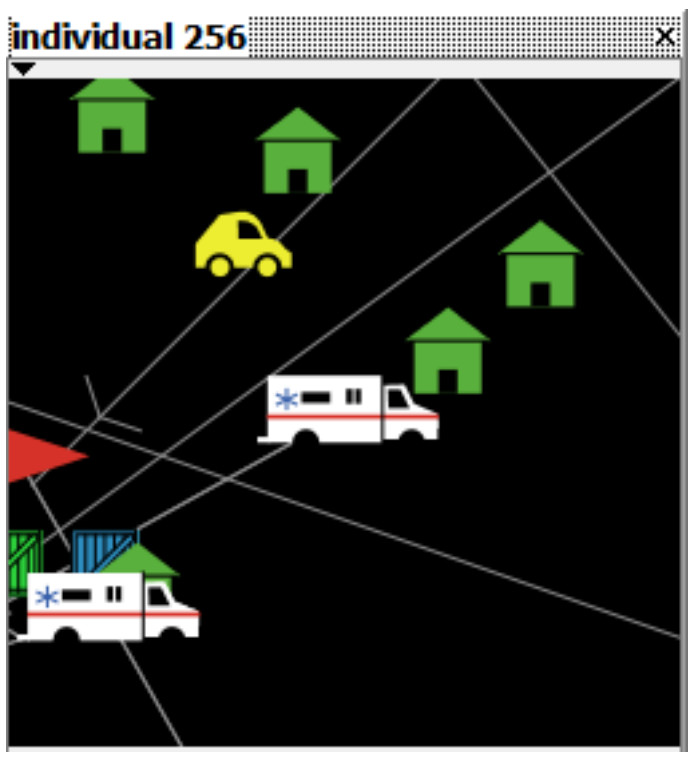

(b) Ambulance transporting the patient.

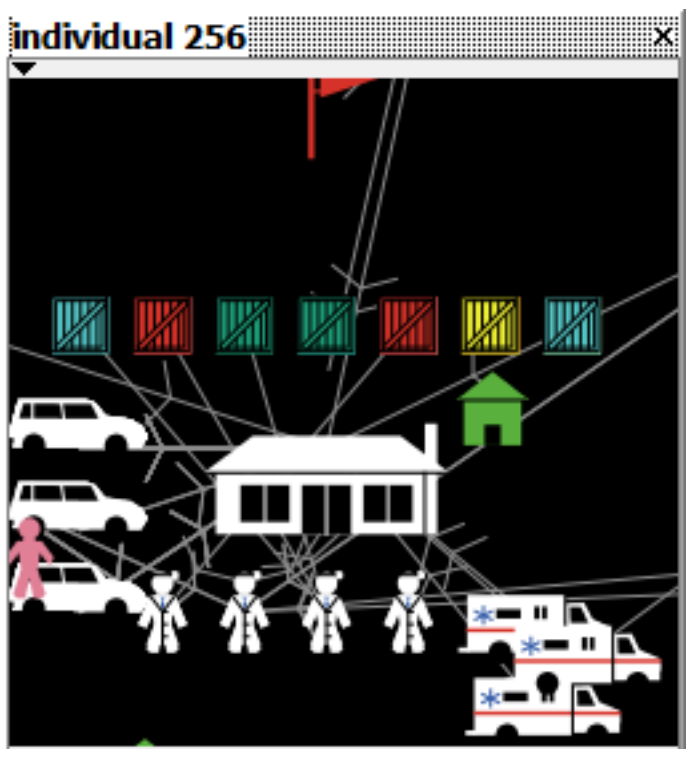

(c) The arrival of patient in hospital.

Figure 4.16: A Netlogo screenshot of phases that "individual 256" goes through in case of "health-care" events.

positive case it replies to the exception by assigning a Doctor and an Ambulance (if necessary) to deal with the given condition of the patient. If it happens that the given Local Hospital is out of available resources capable of handling the given situation, or in case the condition of the patient is complex and the hospital isn't able to cope with it, an exception towards "Regional hospitals" level is triggered. The "Regional hospital" will be set in charge of finding another hospital in possession of missing resources. With this setup the request will traverse the agents spread throughout the layers of FSO organization, until the needed resources are found. This process is manifested by the establishment of a new SON, or a temporary SoC, coupled of shared resources from distinct SoCs, and it is the key feature behind FSOs effectiveness. This represents a case where Mutualistic Relationship is established among members of distinct communities. A graphical representation of the phases a patient goes through, before the start of its treatment is presented in Fig. 4.16

Fig. 4.17 shows an extracted command log from the simulator representing the message flow between the agents. Fig. 4.17a) shows a trivial case of SON, where the received exception, is handled within the local hospital SoC, without a need for additional resources. We call this case an Infra-Community Cooperation. Fig. 4.17b) represents a more complex case, where resources need to be shared between two local hospitals, and this represents the Inter-community cooperation case.

```
    ......
(individual 254): "Exception: I am having health issues, I am in need of urgent health care!"
(people's_coordinator 159): " L2 exception: health care services required for (individual 254)"
(l2_emergency_response_coordinator 158): "(people's_coordinator 159) direct the patient to (localhospital 13)"
(localhospital 13): "(ambulance 22) assigned to provide help for (individual 254)"
    ......
(doctor 85): "Diagnosing patient: (individual 254)"
(doctor 85): "[(medicalappliance 114) (medicalappliance 119) (medicalappliance 129)] will be used to cure (individual 254) for 341 time units."
    ......
(individual 254): "Finished treatment at (localhospital 13)"
    ......
```

(a) Infra-Community.

```
    ......
(individual 271): "Exception: I am having health issues, I am in need of urgent health care!"
(people's_coordinator 159): " L2 exception: health care services required for (individual 271)"
(l2_emergency_response_coordinator 158): "(people's_coordinator 159) direct the patients to (localhospital 13)"
(localhospital 13): "(ambulance 16) assigned to provide help for (individual 271)"
    .......
(doctor 94): "Diagnosing patient: (individual 271)"
(doctor 94): "Exception: 1 medical appliances for curing (individual 271) from condition: 9 missing !!!"
(localhospital 13): " L2 exception:(doctor 94) misses medical appliances to be used for curing condition: 9"
(regionalhospital 9): "(localhospital 10) will offer appliance to (doctor 94) of (localhospital 13)"
(localhospital 10): "(medicalappliance 125) will be used by (doctor 94)"
(localhospital 10): "(transportvan 38) will transport the (medicalappliance 125) to (localhospital 13)"
    ........
(doctor 94): "[(medicalappliance 107) (medicalappliance 113) (medicalappliance 125) (medicalappliance 150)] will be used to cure (individual 271) for 368 time units."
    ........
(individual 271): "Finished treatment at (localhospital 13)"
    ........
```

(b) Inter-community cooperation.

Figure 4.17: Netlogo screenshot of phases that "individual 256" goes through in case of "health-care service" events.

In our simulation model we define three types of exceptions for local hospitals, namely:

- lack of doctors capable of treating a particular type of patients disease
- no ambulances available, to transport patients with severe condition
- medical appliances needed to treat the patient missing.

In this way, by triggering exceptions whenever resources are missing within a particular hospital an interacting channel between hospitals is established, allowing them to function as a greater organization, where resources are shared accordingly with an optimized usage.

The sequence diagrams that depict the flow of messages in cases where the exception is handled within local hospital, local hospital lacks of ambulances for transporting the Individual, and local hospital lacks of doctors capable of curing the Individual are depicted in Figures 4.18,4.19, 4.20 respectively.

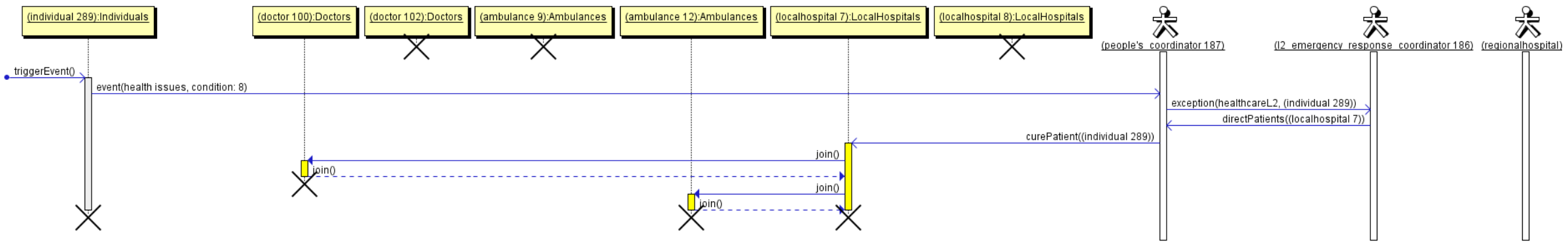

Figure 4.18: A sequence diagram of the flow of messages in the case where the main Local Hospital handles the "healthcare" services exception.

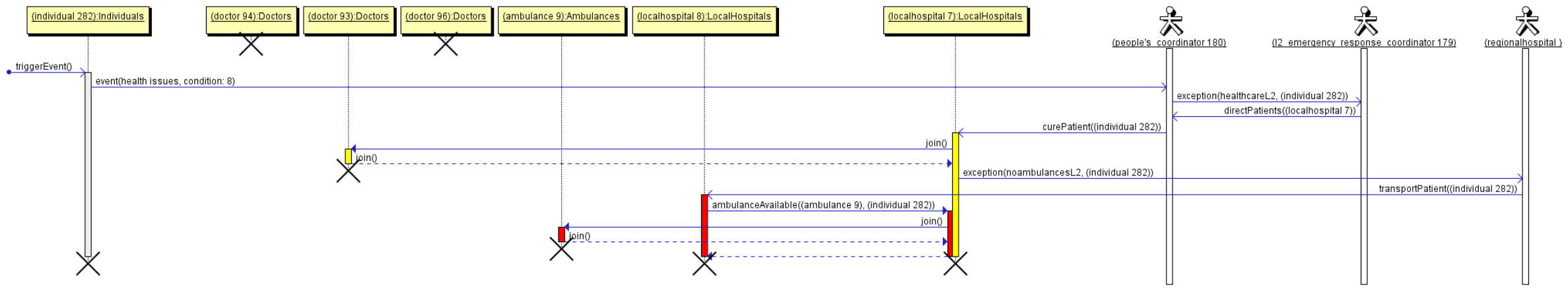

Figure 4.19: A sequence diagram of the flow of messages in the case where "noambulances" exception is handled.

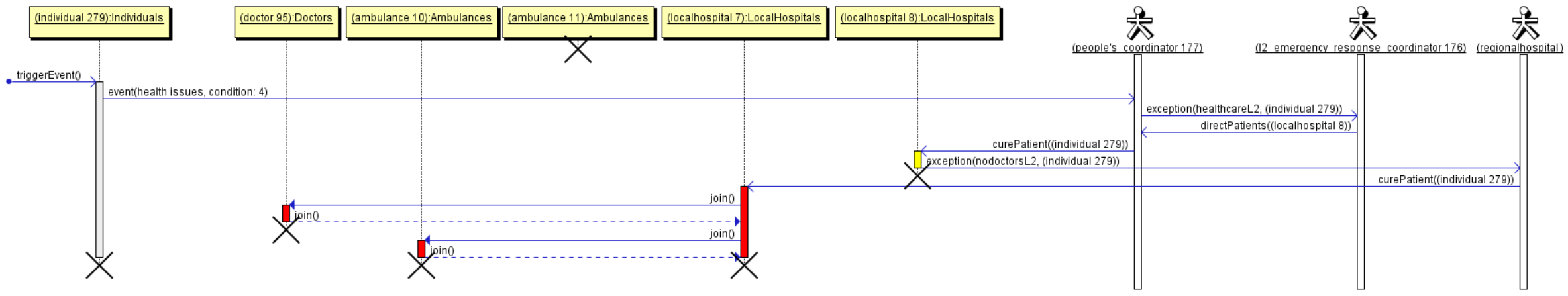

Figure 4.20: A sequence diagram of the flow of messages in the case where "nodoctors" exception is handled

Another scenario considered is the Perfect Oracle. Here the Individuals are assumed to have perfect knowledge regarding the hospitals and services offered. However in this setup there is no resource sharing between the hospitals. To simulate the communication between Individuals and Hospitals the FSO "infrastructure" is used.

A typical scenario would be one where the Individual in need of health-care services reaches the hospital and once the doctor starts the treatment, it finds out that there are no medical appliances available needed for treatment. In this case since the hospitals don't share the resources, the patients request is disregarded and it is directed to another hospital. The process is repeated until the Individual reaches the appropriate hospital that offers treatment, otherwise the request is invalidated.

The health care service request is divided in two phases, the Querying time which represents the time form the moment an Individual is in need of treatment until the moment when all the needed resources are allocated, and the second phase which is the Treatment time and it is the time it takes for a patient to get cured from the moment when the treatment begins. In our experiments we are interested in the former as it represents the measurement that helps differentiating the efficiency of the approaches. The aim of simulations is to set a comparison point between the different approaches and highlight the efficiency of FSO, when dealing with complex situations involving agents generating requests of various natures. The experimental results and a comparison between the introduced approaches is discussed in the next section.

### 4.6.4 Results

In this section we evaluate the benefits of FSO with respect to Perfect oracle and Traditional organization. The experiment outcomes are generated from the NETLogo simulations. The simulations are run for all three scenarios for 3000 "ticks", and with a fixed number of community members. There are 4 Local hospitals defined, as well as 15 Doctors, 8 Ambulances, and 70 Medical appliances. The Doctors, Ambulances and Medical appliances are distributed throughout the 4 hospitals according to the NetLogo random function, generated by a deterministic process. This means that using the same random "seed" the experiments can be reproduced. This approach was useful in our experiments since we wanted to have the same distribution of resources for all three models, in order to have an equitable comparison.

To evaluate the model we use two main metrics, namely the Average Querying Time, and the Number of Patients which could not receive the treatment within the defined threshold, and thus die. The Average Querying time represents the average time an Individual in need for "health care" services has to wait until he receives the necessary treatment. We ran the experiments for each scenario with three different thresholds, as depicted in Fig. 4.21, 4.22. For each specified threshold the simulation is ran 5 times, with a distinct number of Individuals, starting with 60 until 140, with an increasing step of 20. Additionally for the FSO we use the SON metric, which describes the number of times a solution to the given situation is solved as a result of inter-community cooperation. This number also represents the usage of communities social energy. Contrary to FSO, for the Traditional Organization where there is no cooperation between the Hospitals, we measure the number of failures a patient faces until the matching hospital for treatment is found.

Figure 4.21 shows a comparison of the number of treated patients, and the number of patients died for each scenario. For example, in the simulation experiment with a Threshold of 150, for 140 Individuals the following results are archived:

**FSO** 189 patients are treated, out of which 56 have died,

**Perfect Oracle** 177 treated, 71 died,

**Traditional organization** 175 treated 93 died.

Not only the number of patients treated in FSO is higher, but also the number of deaths is lower. It can be observed that in all cases FSO performs better, followed by Perfect Oracle, and Traditional Organization. In the FSO case the number of patients that fail to receive the treatment is always lower, due to the sharing of resources between the hospitals, while for Perfect Oracle this number is higher because the patient knows at which hospital he should go, but if the hospital lacks necessary medical appliances it doesn't borrow them from other hospitals, it redirects the patient instead. In the traditional organization we have no interaction between hospitals neither. However, in this case the patient chooses the hospitals randomly, without any knowledge about the available doctors, ambulances or medical appliances. This poor level of information leads to delays in resource discovery, thus the number of patients that fail to receive treatment is the highest.

The fast service response-time of FSO, also allows for a higher number of patients to be treated for the same number of simulation cycles. This is observed in the experiment results too.

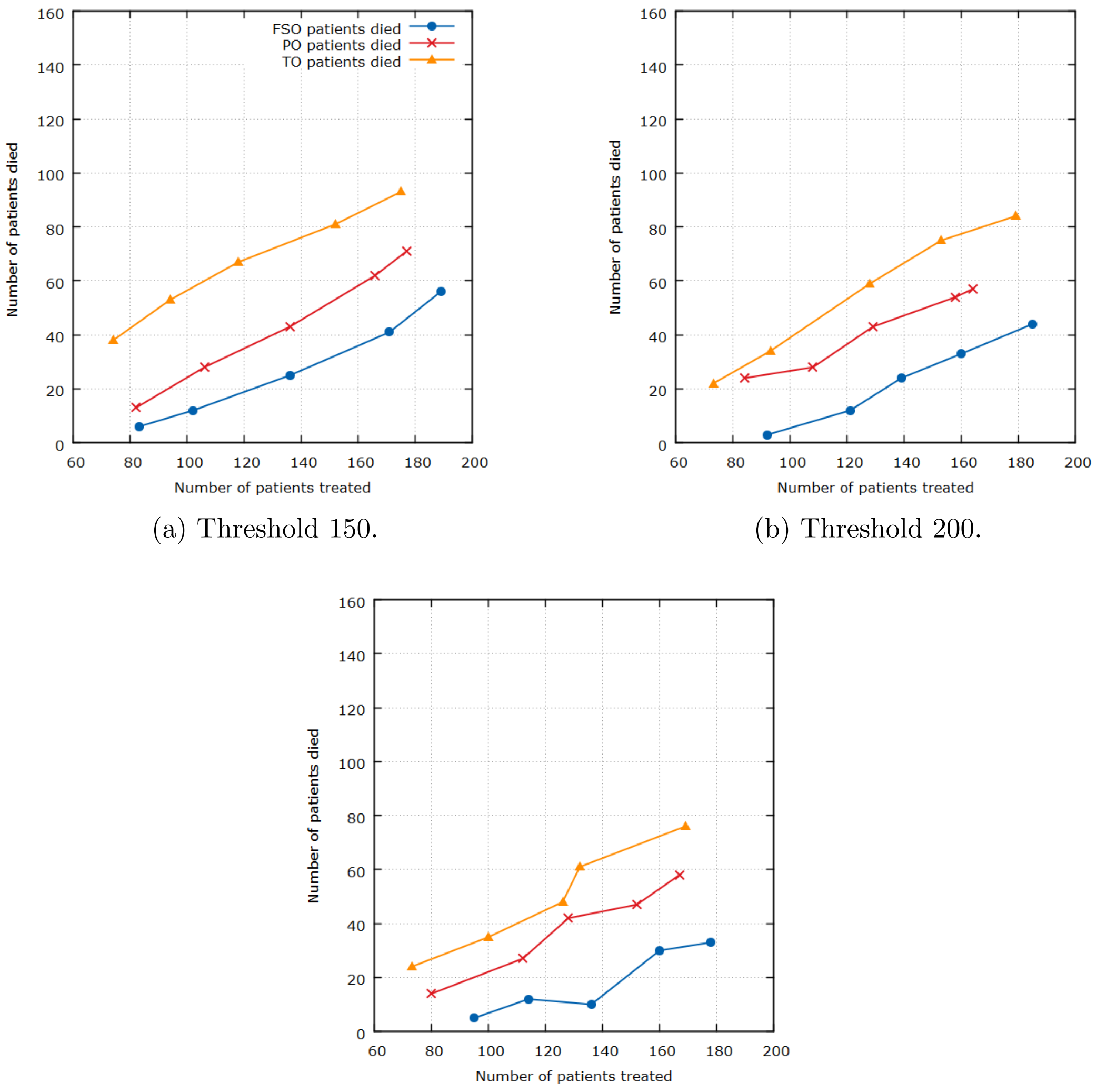

(a) Threshold 150.

(b) Threshold 200.

(c) Threshold 250.

Figure 4.21: Comparison of patients treated/died ratio for each of the three scenarios for various threshold values.

Another observation if we compare the scenarios from the three graphs of Figure 4.21, is that with an increase of the Threshold the number of patients that die decreases, same as the number of patients treated. This because the request for “health care” services has more time to receive a response, thus less requests are invalidated. The number of patients treated also decreases, as the number of generated cases decreases slightly, because of the longer period of service reception.

Fig. 4.22 depicts the performance of all three models with respect to Average Querying time. If we analyze the querying time results, again we see that FSO outperforms the other two models. Such outcome comes as a result of exceptions usage, which allow for fast inter-community communication and resource allocation. It can be observed that by decreasing the number of Individuals, the FSO improves drastically, and the number of patients that fail to receive the treatment decreases two to three times.

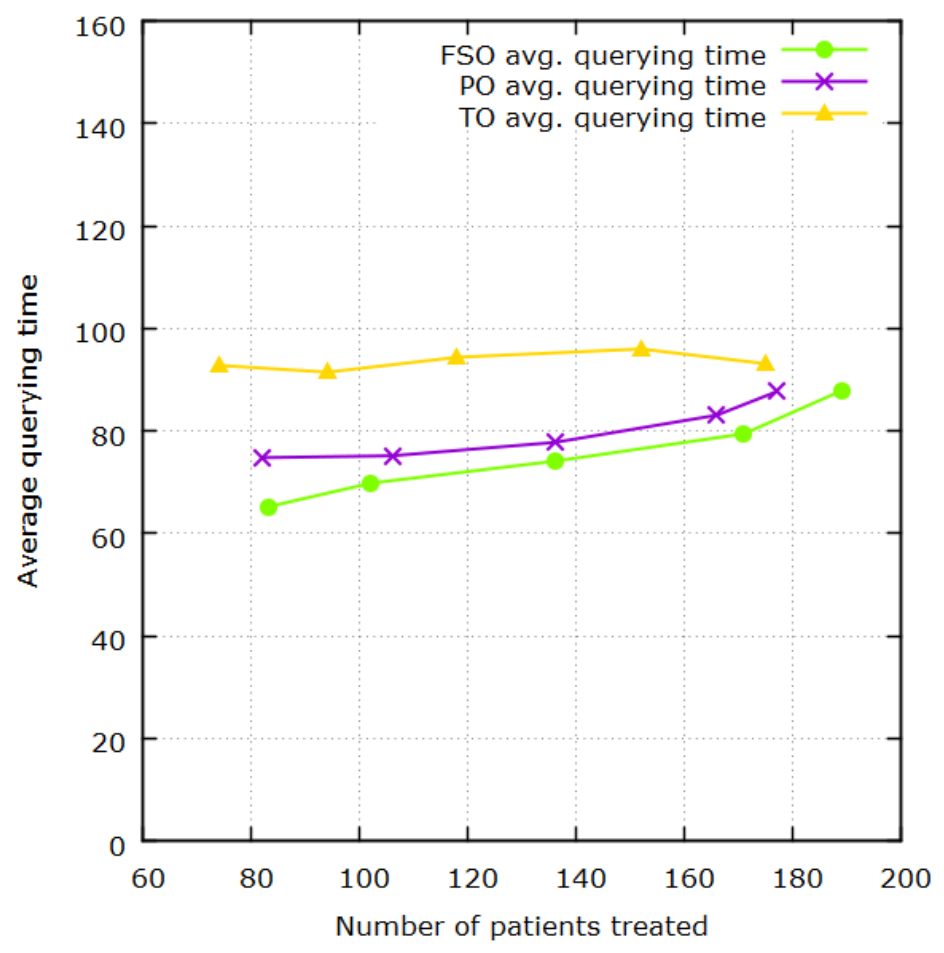

(a) Threshold 150.

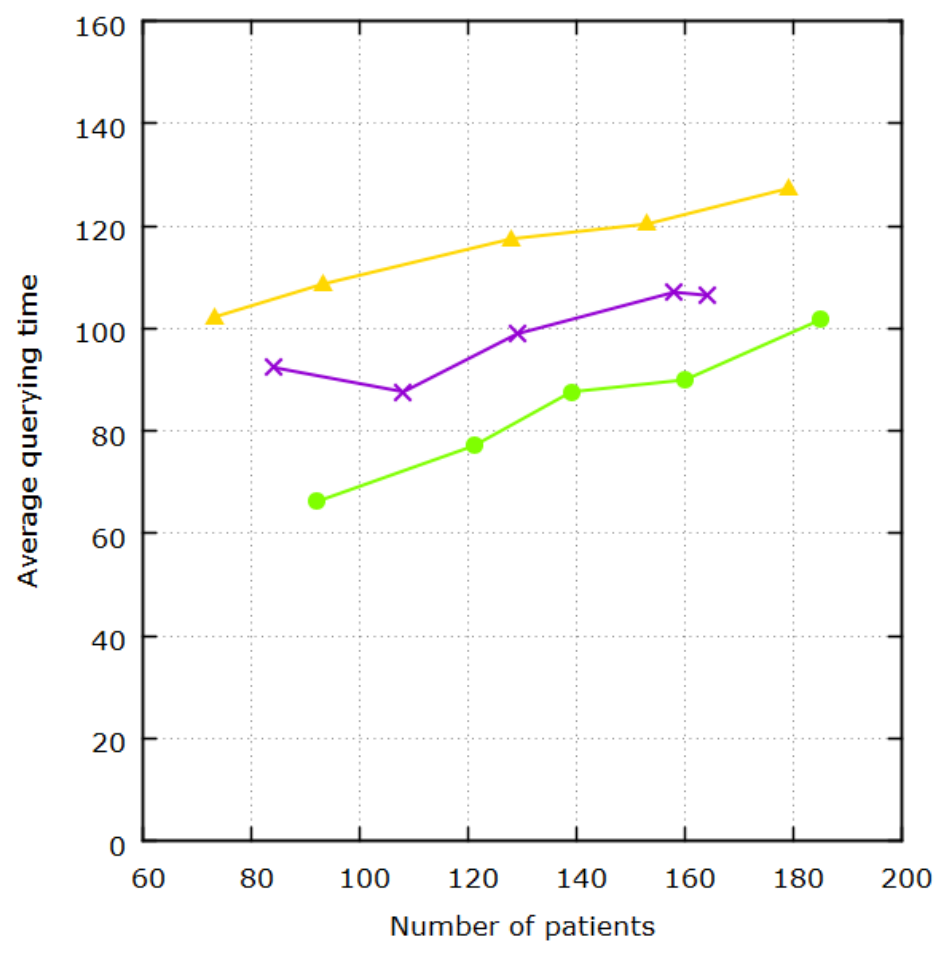

(b) Threshold 200.

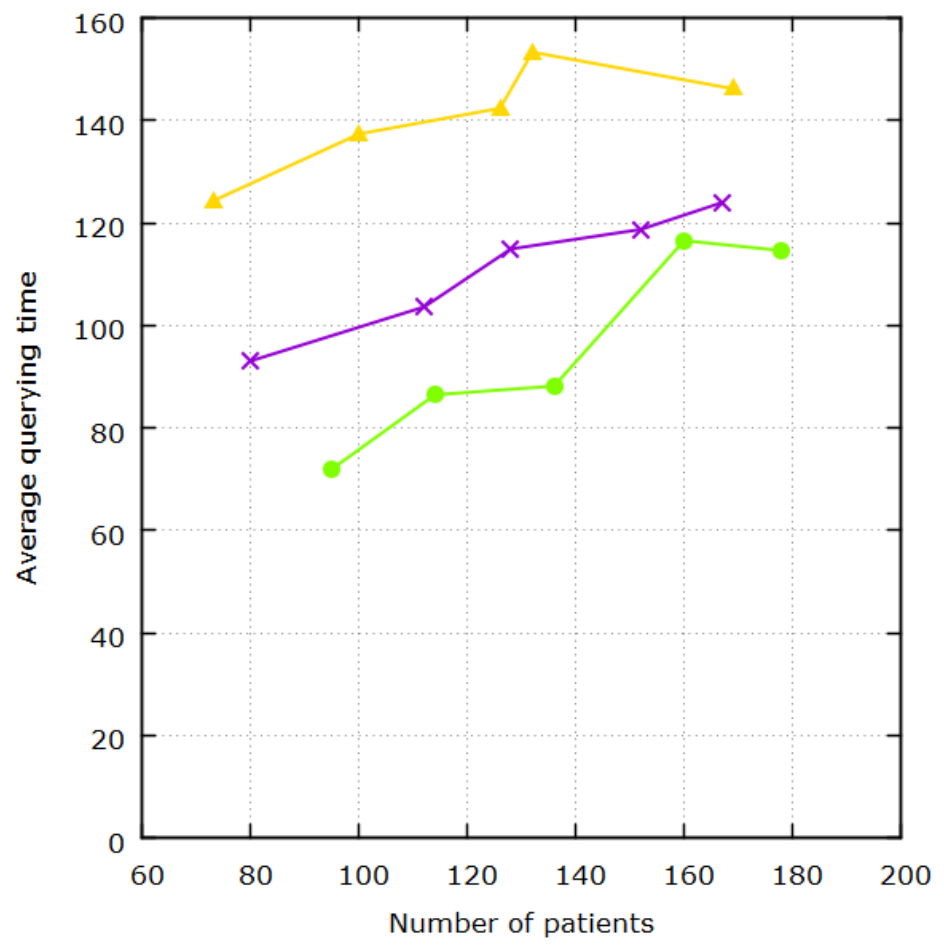

(c) Threshold 250.

Figure 4.22: Average Querying time for each of the three scenarios for various threshold values.

The SON metric used for FSO ranges from around 100 up to 220 depending on the number of Individuals participating in the simulation. The higher the number of Individuals, the higher the SON metric. This due to the increase of number of collaborations between communities. On the other hand the number of service failures in the Traditional Organization ranges from around 50 up to 200. By increasing the number of Individuals the number of failures increases too. This because the number of times a patient fails to reach the appropriate hospital increases with a higher number of Individuals.

# Chapter 5

# Second model

## 5.1 Overview

Falls present the major public health problem among elderly persons. According to [19] falls are the "most significant cause of injury for elderly person" —more specifically "the most serious life-threatening events that can occur" in the 65+ age group [20]. The key factor that determines the severity of a fall is the time until the treatment is delivered [22],[20].
The number of systems aimed at fall detection has had a dramatic increase over the recent years. Most of them are based on devices such as accelerometers, gyroscopes, or other sensors. A major limitation of such systems lies on the difficulties of the algorithms and their implementations to provide a reliable assessment for a fall under real life conditions. Other approaches are oriented towards smart-phones and their rich set of embedded sensors. [23],[24],[25]. Despite the continuing technological progress, it is still very difficult to find a monitoring system that is able to determine whether a person has actually fallen or whether, for example, he or she has changed his position very quickly. A possible improvement might come by the use of redundancy. For example two accelerometers of different technology and design are coupled together [19]. As a result we would have an improved sensitivity. However, this solution comes with an impact on the social costs spent for the verification of the false alarms.

There are two possible orientations when one tries to approach this issue. One way may be to try and improve the current ITC monitoring systems. There have been numerous research actions, with the intention to improve both sensitivity and specificity parameters, and yet no standardized solution has been found [22]. Another possibility it to search for alternative solutions. Here we discuss one way to do so. Our proposal is to go beyond the purely technological solution and involve the humans in the system. Consequently, a socio-technological system is introduced and analyzed by means of the provided multi-agent simulation model. By the use of the organizational structure of the FSO we show that the appointment of fall verification tasks to a group of human agents allows to overcome, to some extent, the limitations and inefficiencies of purely ITC-based solutions.

In what follows we briefly introduce a number of key measurements that will help us evaluate the outcomes of our simulation model:

**False positive ratio** False Positive (FP) ratio or false alarm rate is the probability of concluding that an event occurred, whereas this is not the case. The alarm is fired, although the event did not take place.

**Specificity** Specificity is the probability that the event did not take place and the system did not fire. Specificity is equal to $1 - \text{FP}$.

**False Negative ratio** False Negative (FN) ratio is the probability that the system indicates that an event has not occurred, when in fact it took place. It can also be described as a missed alarm.

**Sensitivity** Sensitivity is the probability that the event took place and the system did fire. Sensitivity is equal to $1 - \text{FN}$.

In the next section we will describe the main characteristics of the implemented simulation model.

## 5.2 Simulation Model

Again for the setup of the simulation model we make use of the NetLogo environment. The “virtual world” is now coupled of passive agents named the Elderly persons. Each of these agents is associated with one or two monitoring devices, depending on the simulation scenario. The Elderly agents are positioned in a static position in the simulation area, as an assumption of their living area (e.g. home). There is one type of event that occurs, namely the fall of Elderly agents or the "fall event". The two monitoring devices, (e.g. accelerometer, gyroscope) are in charge of identifying such events, and raising alarms towards the Community Agents. However, these devices aren't always accurate identifying the "fall event". The alarms are collected by the first level Community agent or the so-called "Middleware 1", which then forwards them to the next level agent, namely the "Middleware 2". "Middleware 2" is responsible for allocation of available Informal Carers for visiting the Elderly persons whose alarm has been triggered. The Informal Carers are agents that wander around the simulation area until they receive messages from the Community agent that asks them to go towards a certain location, where an Elder person is residing. Once at the Elder agent's site, the Informal carer verifies if the triggered alarm was true.

At the same time the alarm message is forwarded to the "Middleware 3" community agent, which represents the Hospital coupled of moving agents or Ambulances, and Professional carers or Doctors. In cases where the triggered alarm is True, the intervention of Ambulances is required to transport the Elderly person to the hospital, for receiving the necessary treatment. In what follows a formal definition of the above described agents is given.

**Elderly Agents.** Elderly Agents (EA) are agents representing elderly or impaired persons residing in their house where their condition is monitored by Device Agents. In case of true fall detection the EA take the role of patient, thus professional care is requested.

**Professional Carers.** Professional Carers (PC) represent agents able to supply certified healthcare services to Elderly agents. Such agents are institutional, and provide professional service (e.g the doctors of a hospital).

**Informal Carers and Verification.** Informal Carers (IC) are mobile agents capable of providing non-professional services. It is assumed that IC can pay a visit to an EA and report whether the EA has truly experienced a fall or not. This is called in what follows *verification*.

**Device Agents.** Device Agents (DA) are simple devices (e.g. accelerometers, mo-

tion sensors, cameras etc.) that are able to provide monitoring services. DA's are attached to EA agents, thus monitor their condition. A DA triggers an alarm when it ascertains, with a certain probability, that an EA has fallen. Another example of DA is a motion sensor stating that an observed EA is not moving.

DA are characterized by a non-zero probability of false positives and false negatives. Thus for instance in some cases an accelerometer may detect a fall when this is not true, while in some other cases the accelerometer may not detect a true case of fall.

**Mobility Agents.** Mobility Agents (MA) are agents able to provide mobility or transportation services to other agents. In our case such agents are represented by ambulances.

**Community Agents.** Community Agents (CA) are agents managing a "circle" of other agents. A detailed description of such agents was provided in Chapter 3.

A visualization of the simulation area containing the initialized agents can be seen in Figure 5.1. The green "person" shape represents the EA agents, while the connected devices represent the DAs. The ICs can be seen close to their cars, used for transportation. The white building represents the hospital, and connected to it we have the MAs and PCs.

In terms of the FSO approach, the core unit of the given organizational structure is the iSoC coupled of an EA, and corresponding DAs. Further all iSoCs form a Service-oriented Community (SoC) coordinated by the Level 1 CA. The IC agents reside within the Level 2 CA forming another SoC, similarly to the Level 3 CA coupled of MA and PC. The agents associated with a CA may send it notifications. Whenever there is a need for communication between the CAs of distinct levels, the "exception" messages are triggered. It is assumed that these messages are transmitted reliably and instantaneously. The overall FSO structure is depicted in Figure 5.2

In order to quantify the results of the various simulation scenarios described in the upcoming section, we introduce the following metrics:

**Social Cost.** We define Social Cost (SC) as the number of cycles that a social

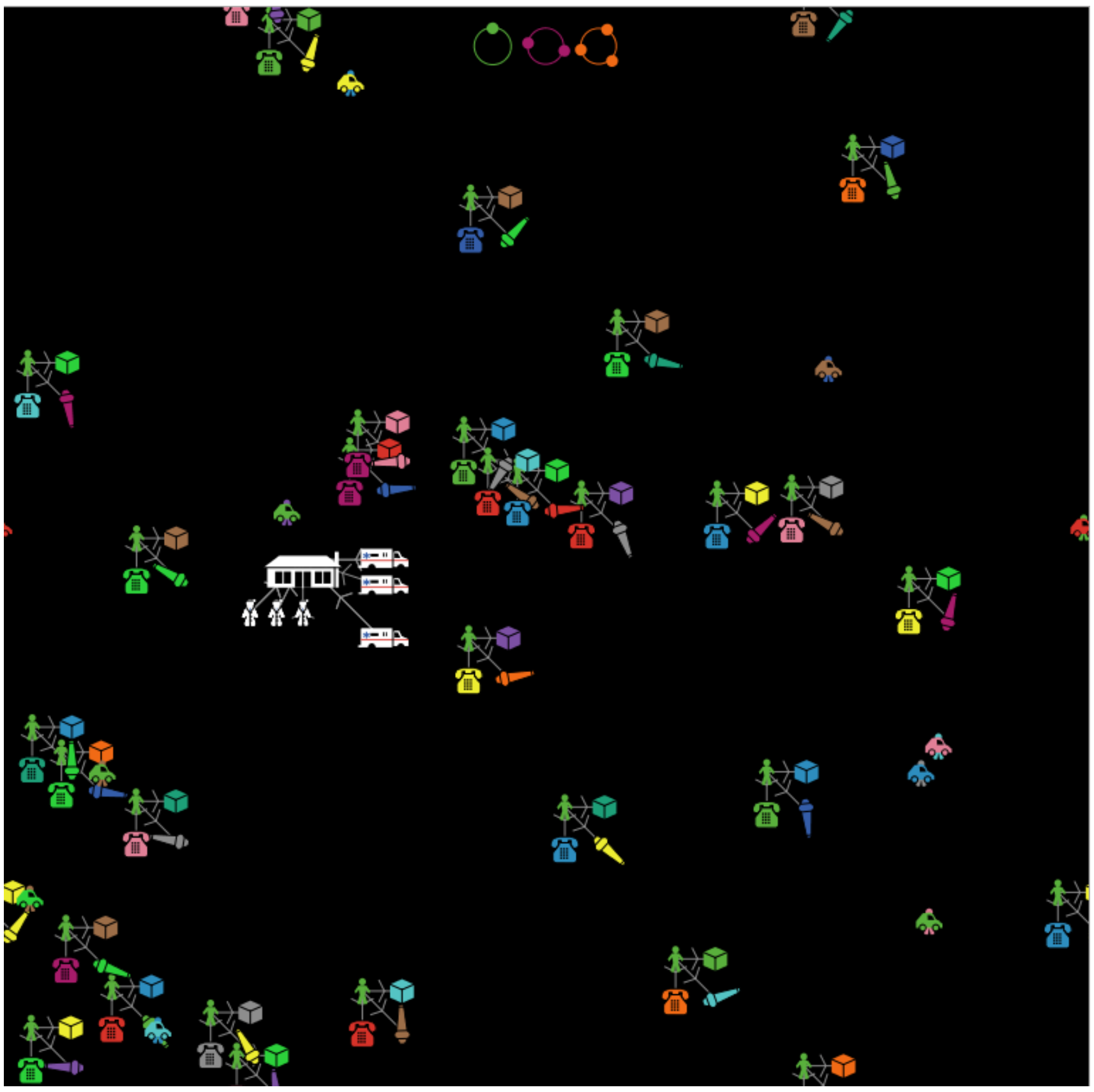

Figure 5.1: NetLogo screenshot of the simulation area with initialized agents.

resource is used to intervene either for a true alarm or for a false positive (FP).

**Cumulative Social Cost.** Cumulative Social Cost (CSC) is defined as the overall number of cycles used by a community to deal with true alarms and FP's throughout a (simulated) time interval $T$.

**Waiting Time.** Waiting Time (WT) is the number of cycles elapsed from the moment an alarm is triggered, being it true or false.

**Cumulative Waiting Time.** (CWT) is the overall number of cycles elapsed from the moment an alarm is triggered, being it true or false. It is the sum of all the individual WT's occurred throughout a simulation run.

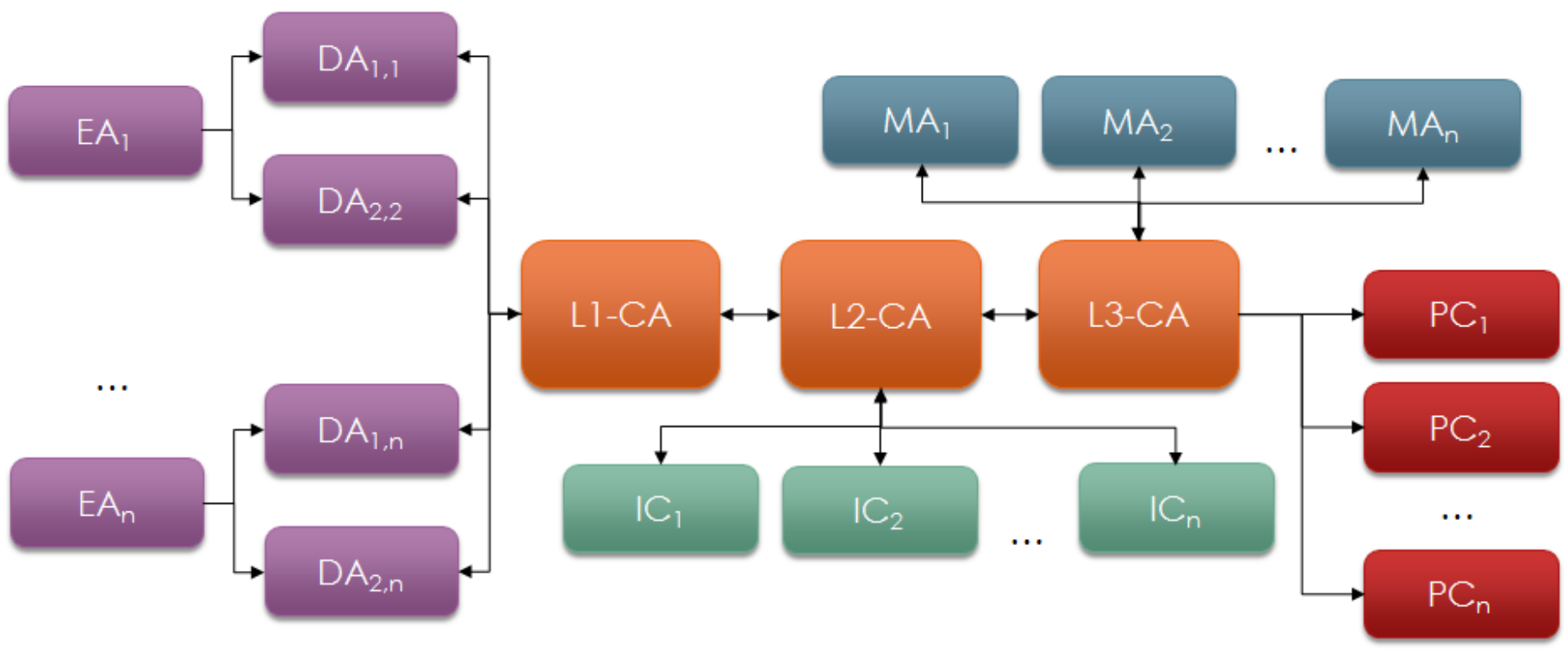

Figure 5.2: A representation of the FSO structure of the simulation model.

## 5.3 Simulation Scenarios

Based on the participation of agents on the simulation, we distinguish three main scenarios:

$S_1$ : Only accelerometers. In this scenario we use only one DA to monitor the state of EA. In addition, there are no ICs defined meaning that there is no Level 2 CA. This represents the simplest scenario, where once an alarm is triggered, the message is directly transmitted to the Level 3 CA or the hospital. The CA is responsible for allocating the MAs to pay a visit to the EA. In case treatment needed the EA is transported to the hospital, where a PC is assigned to handle the case at hand. We measure FP rate, FN rate, Sensitivity, Specificity, CSC and NCSC during a given $T$. CSC refers to total number of cycles in which MAs are serving the EAs.

$S_2$ : Accelerometers combined with a second sensor (camera or gyroscope [19]). This scenario is the same as $S_1$, with a difference in the DA number. Here we use two DAs, instead of one. This affects the FP, FN rate, Sensitivity and Specificity. We measure the same as in $S_1$.

$S_3$ : One or two DAs and Informal carers [21, 26]. In these scenarios the complete FSO structure as described earlier is used. $S_3$ with one DA represents the $S_1$ with added ICs, while $S_3$ with two DAs is the $S_2$ with added ICs. We measure as in $S_1$ and $S_2$, with various numbers of ICs. Here as described in the results section we observe a better usage of Social resources.

We consider the following configuration of agents for the given scenarios:

- 30 EA's residing in their houses. House locations are assigned in cells chosen pseudo-randomly in the virtual world.
- In scenario $S_1$, 1 DA (an accelerometer) is deployed with each EA. In scenario $S_2$, an additional DA (e.g., a gyroscope) is added for each EA.
- 1 hospital (Level 3 CA), located in a cell chosen as described above.
- 6 PC's and 5 MA's (ambulances) are located with the hospital.

Scenario $S_3$ extends scenarios $S_1$ and $S_2$ with the addition of IC's. We perform simulations with an increasing step of 5 IC's from 0 to 40. The IC's are initially assigned to pseudo-randomly chosen cells and then wander randomly within the boundaries of the virtual world. The IC's receive requests for verification from Level 2 CA. Once this happens, they quit pseudo-random wandering and move from their current position to the verification place.

The major steps in the general scheme of execution for the three scenarios are represented as pseudo-code in Tables 5.1, 5.2, 5.3, 5.4, and 5.5.

Table 5.1 shows the pseudo-code of the procedure of DA agents. It represents a Simple "reflex": if a fall is suspected for the corresponding EA, an alarm is triggered towards the local CA. The alarm can be raised only when the EA is residing home. The only moment when EAs might not be there are the cases when they had a

```
Procedure DA// Code of the DA agents.
            // Reaction to false or true alarms.
begin
    When ((this.Fired == TRUE) and (EA.NonFallingPeriod == 0) )
        this.Send(CA, ALARM, this);
    EndWhen
end
```

Table 5.1: Pseudo-code of the DA agents.

```
Procedure L1-CA // Code of the Level 1 CA, reaction to alarm message
begin
    When (∃d ∈ DA (this.AlarmReceived(d) == TRUE))
        Parallel
    // In parallel, raise exceptions to Level2-CA and Level3-CA
            Do exceptionEAs[] ← this.Exception(d, ALARM);
            this.Send(exceptionEAs, L2-CA);
            this.Send(exceptionEAs, L3-CA); EndDo; EndParallel
    EndWhen
end
Procedure L2-CA // Code of the Level 2 CA, allocation of IC agents.
begin
    When (∃i ∈ IC (i.getAvailable() == TRUE))
        // Assign IC to visit EA for which alarm is raised
            forEach d in exceptionEAs;
                Do this.assign(i, d); i.setAvailable(FALSE); EndDo
            EndForEach
    EndWhen
end
Procedure L3-CA // Code of the Level 3 CA, allocation of MA and PC agents.
begin
    When (∃p ∈ PC (p.getAvailable() == TRUE)
                && ∃m ∈ MA (m.getAvailable() == TRUE) )
        // Assign MA to visit EA, and a PC in case intervention needed
            forEach d in exceptionEAs;
                Do this.assign(p, m);
                 p.setAvailable(FALSE); m.setAvailable(FALSE);EndDo
            EndForEach
    EndWhen
end
```

Table 5.2: Pseudo-code of the CA agents.

treatment earlier in hospital and are on their way back. Thus, the Non-falling period is used.

The pseudo-code of the CA agents is presented in Table 5.2. It can be observed that procedure L1-CA handles the alarms received from DAs. Further it redirects them as exceptions to Level 2 and Level 3 CA. The L2-CA procedure allocates the available ICs to perform the verification, while L3-CA allocates the MA and PC agents based on the received "exception" messages.

The IC's wander aimlessly through the virtual world until a verification request is received. The request is served, possibly canceling an ongoing emergency call. The IC procedure can be seen in Table 5.3.

```
Procedure IC // Code of the IC agents.
             // Reaction to alarm received by L2-CA.
begin
    When (this.AlarmReceived(L2-CA, d) == TRUE))
        Do
            // The IC agent goes to the cell the alarm came from
            this.GoTo(location(d));
            // Once there, it verifies whether
            // it was a case of FP (false positive)
            If (this.Verification() == FALSE)
            // If that is the case, it informs L3-CA to cancel the
            // emergency call, and free the PC and MA resources,
            // thus reducing the waste in social energy
                this.Send(L3-CA, CANCEL_ALARM, d); EndIf
        EndDo
    // if no alarm is received, the IC agent just wanders pseudo-randomly
    Otherwise this.Wander();
    EndWhen
end
```

Table 5.3: Pseudo-code of the IC agents.

The MA agents wait for emergency calls. When one such call arrives, the MA agents (ambulances) start moving towards the EA for which the alarm was raised. Reaching that location is a time- and resource-consuming action. Said time and resources are wasted in the face of a FP event. In the case of Scenario $S_3$ the loss is reduced if an IC agent reaches the location earlier than the MA and verifies the occurrence of an FP. In such a case, the IC cancels the call, thus sparing some social energy. The pseudo-code of MA agents can be seen in Table 5.4.

Finally the PC procedure is presented in Table 5.5. The PC's check for the arrival of the MA carrying the EA agent. Once there they start the treatment. When the treatment process finishes the EA's are freed, and PC's can be assigned to another case.

```
Procedure MA// Code of the MA agents.
            // Reaction to alarm received by the L3-CA.
begin
    When (this.AlarmReceived(L3-CA, d) == TRUE)
        Do
            this.GoTo(location(d));
            // Abort the emergency call if some IC canceled the alarm
            While (this.AlarmCanceled(IC, d) == FALSE);
            // Once at the location of the alarm,
            // verify the situation, if needed bring EA to hospital
            If (this.Verification() == FALSE)
                this.Send(L3-CA, CANCEL_ALARM, d);
                L3-CA.Send(L2-CA, CANCEL_ALARM, d);
            Else
                this.Bring(EA, location(L3-CA)); EndIf
            EndWhile
        EndDo
    EndWhen
end
```

Table 5.4: Pseudo-code of the MA agents.

Each scenario is executed for 10000 simulation cycles (ticks). Every EA experiences a "true fall" with probability $\frac{1}{600}$, while the DA agents trigger false positive alarms with probability $\frac{1}{500}$, and false negatives with probability $\frac{100}{500}$.

We now describe the results obtained in the three above-sketched scenarios.

```
Procedure PC // Code of the PC agents.
             // Treatment of EA, once in hospital premises.
begin
    When (MA.arrived(d) == TRUE))
        Do
            // The PC agent starts treating the EA
            this.startTreatment(d);
            // Once the, EA treatment finishes, it is free to go
            When (d.treated() == TRUE))
            // a non-falling period c is assigned to EA agent
                d.setNonFallingPeriod(c);
            EndWhen
        EndDo
    EndWhen
end
```

Table 5.5: Pseudo-code of the PC agents.

## 5.4 Results

We start by running the $S_1$ and $S_2$ scenarios without any IC, then for each scenario we run a number of experiments with various number of ICs starting from 0 to 40 ($S_3$ scenario). As mentioned previously there are various measurement metrics used for evaluation. The results of the $S_1$ scenarios are shown in Table 5.6, while for $S_2$ in Table 5.7.

We start by inspecting the FP, FN, Sensitivity, and Specificity parameters. If we compare scenario $S_1$ to $S_2$ without ICs, we can observe that in $S_2$ the number of FPs increases for more than 100 FP units. This due to the addition of a second DA. When there are two types of DA (sensors) per EA defined, the alarm or FP is triggered as a result of an OR operation of the alarm of each DA. Even, if the alarm triggered is true in only one of the sensors we will have a FP. Thus, 3 out of 4 possible combinations will be positive, resulting in a larger number of FPs. Further, by adding the IC agents in the $S_3$ scenarios we see that FP number increases. This because the IC agents confirm the FP alarms in a faster pace. Thus, for the same simulation period of 10000 "ticks" more FP requests are handled (see Figure 5.3). For $S_2$ with ICs the number of FPs fluctuates in the 620 range, while for $S_1$ with ICs around 320. The FN values in $S_1$ are almost twice of those in $S_2$. This because in $S_2$ an AND operation is performed between the FN outcome of DA1 and DA2. As a result the probability of having a FN decreases in $S_2$, and only when we have a FN from both alarms the FN is reported. Comparing these two scenarios with their corresponding $S_3$ cases we see that the addition of ICs doesn't have any effect. This because ICs cannot identify FNs. The graph in Figure 5.4 depicts the FP ratios of $S_1$, and $S_2$ with various number of ICs. The FN ratios are visualized in Figure 5.5.

A larger number of FPs and a smaller number of FNs, provides with a more Sensitive system. This because by triggering more alarms, more cases where the alarm is truly positive are covered. On the other hand there is no difference observed with regard to the Specificity parameter for each scenario. From the results in Table 5.6 and Table 5.7 we see that for $S_1$ with or without volunteers the average Sensitivity is 79.55%, while for $S_2$ it increases to an average of 89.92%. The Specificity values stay the same for all scenarios with an average of around 99%.

We proceed the evaluation with a comparison of the Social costs for $S_1$ and $S_2$,

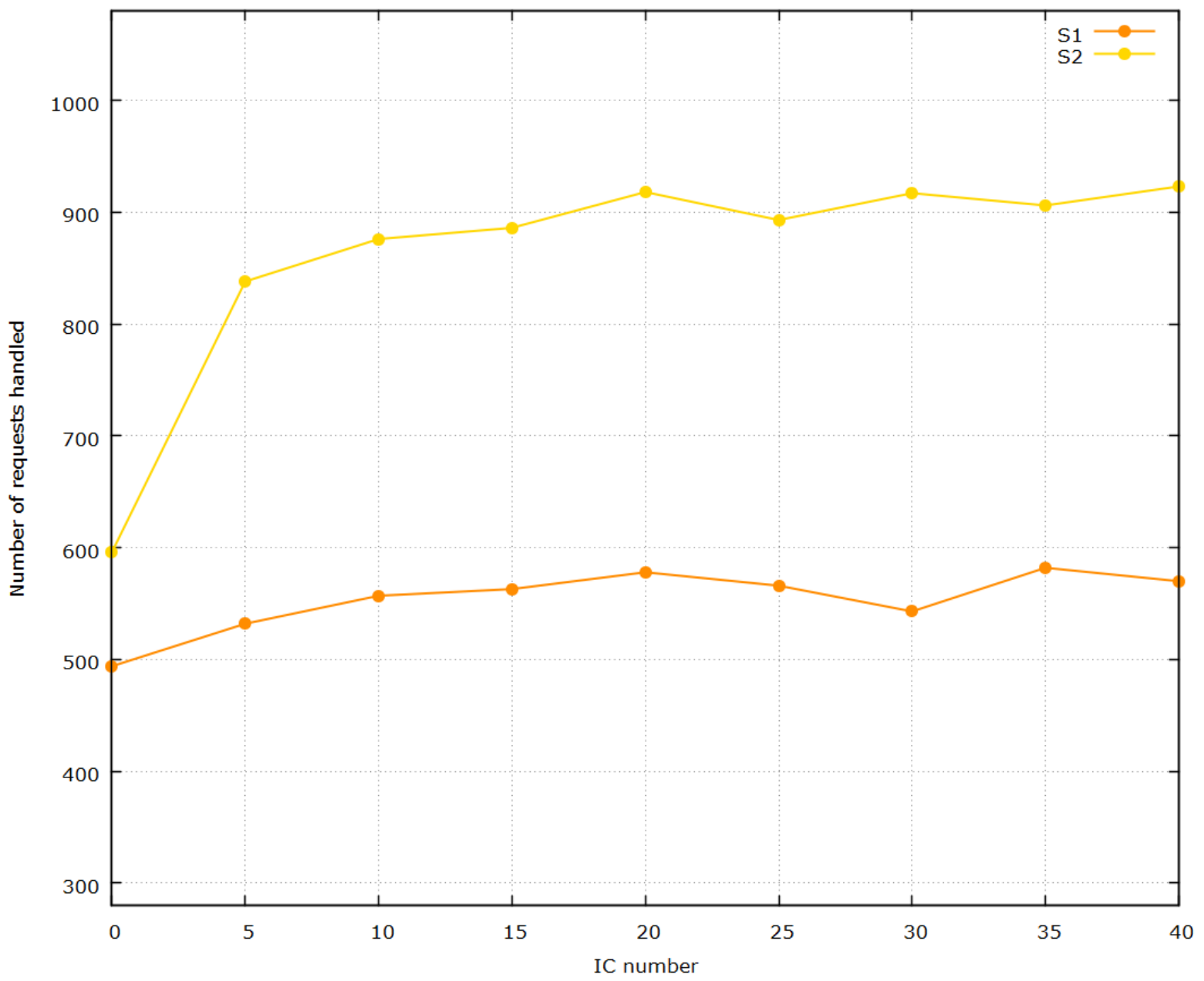

Figure 5.3: Number of alarm requests handled by the system for S1 and S2 with a number of IC's ranging from 0 to 40.

while we continue increasing the number of IC's. Comparing $S_1$ without any IC, with the scenarios where the IC's are added we observe a significant decrease in the cost of MAs. This because the ICs help verifying the condition of the EA agents, thus reducing the cost of MAs. MAs have to intervene only when there has been a true fall of the EA. The same applies for the $S_2$ scenario. Figure 5.6 depicts the relation of average MA cost with the IC number for both scenarios. In the case of S1, the minimum is reached for 20 IC's after which average social cost appears to stabilize. Another metric that shows the impact of ICs in the MA cost is the number of alarm verification's. In all scenarios where IC are present the number of verification's performed by MA is reduced drastically.

An important metric is the average time until a triggered alarm for an EA gets verified, being it true or false. The average waiting time reflects the reaction time of the system for a given event. The best results are obtained with a number of 10 ICs. Additional effort beyond this value produces little improvements. With IC = 10 no significant differences are observed between $S_1$ and $S_2$. A graph showing the average waiting time in report to IC number is given in Figure 5.7.

All in all, we can observe that the involvement of IC agents in the fall detection system using the FSO approach, helps reducing the MA social costs, while at the same time the response time to fall events is fastened.

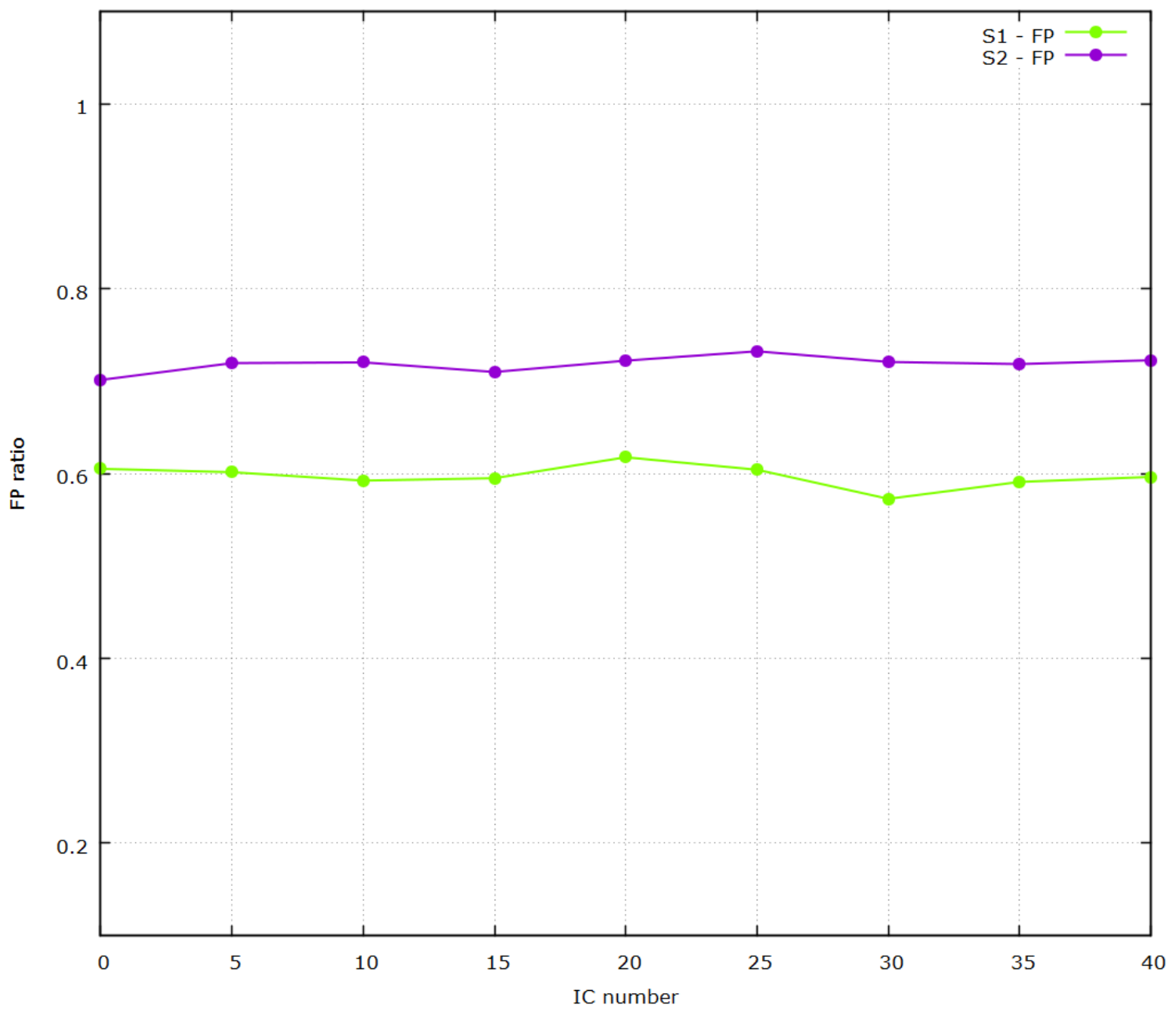

Figure 5.4: FP ratio for S1 and S2 with a number of IC's ranging from 0 to 40.

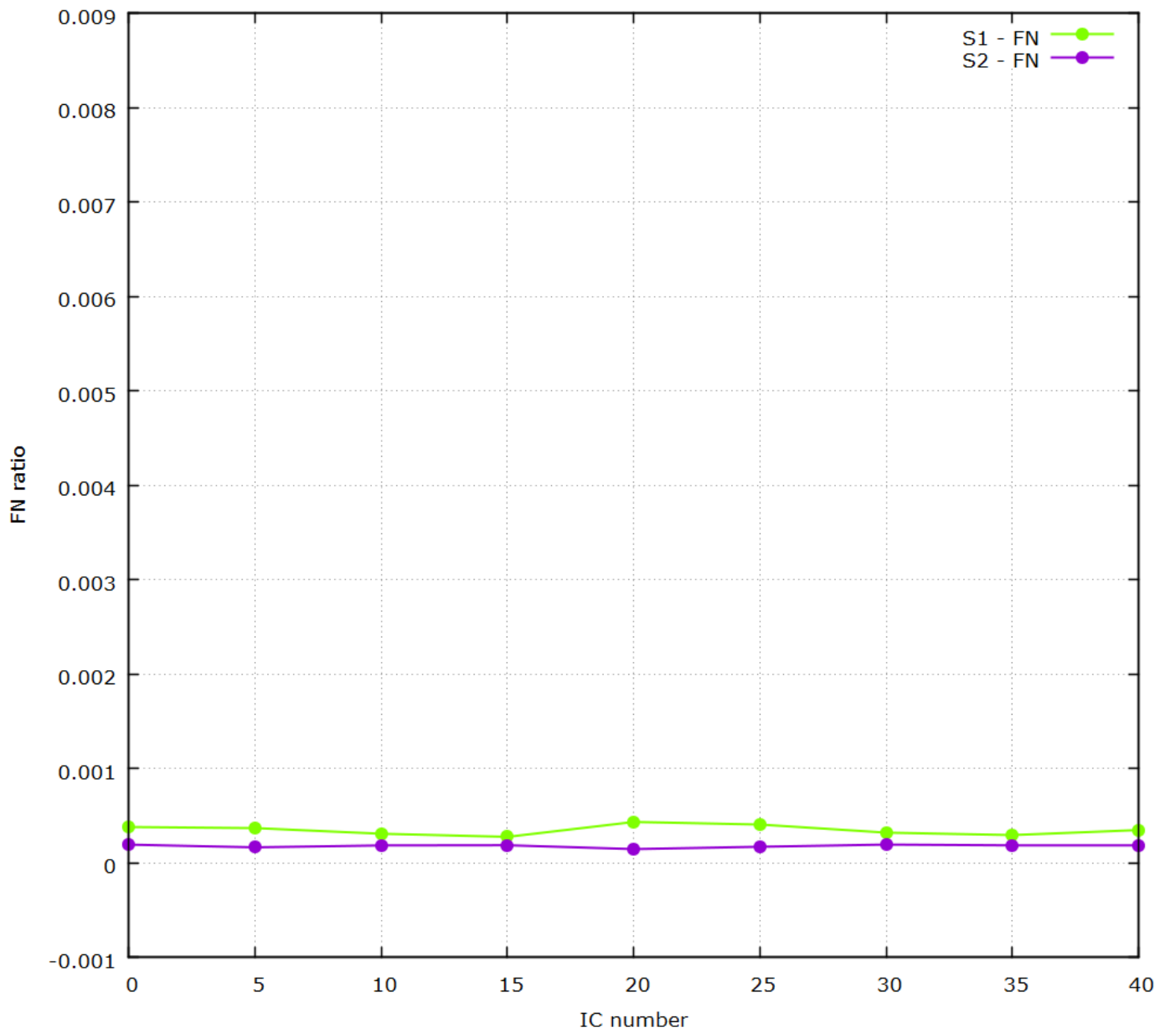

Figure 5.5: FN ratio for S1 and S2 with a number of IC's ranging from 0 to 40.

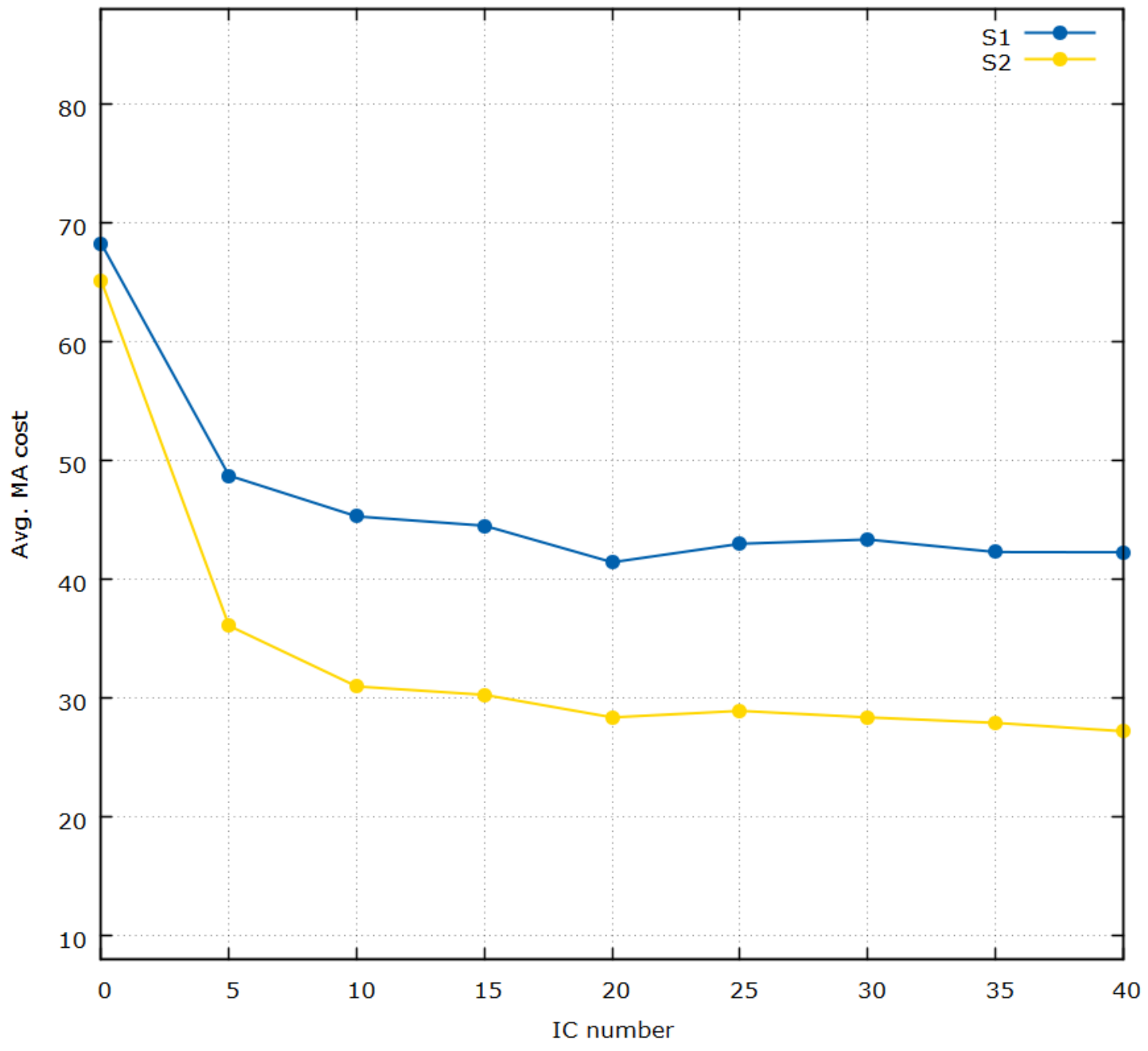

Figure 5.6: Average social costs of S1 and S2 with a number of IC's ranging from 0 to 40.

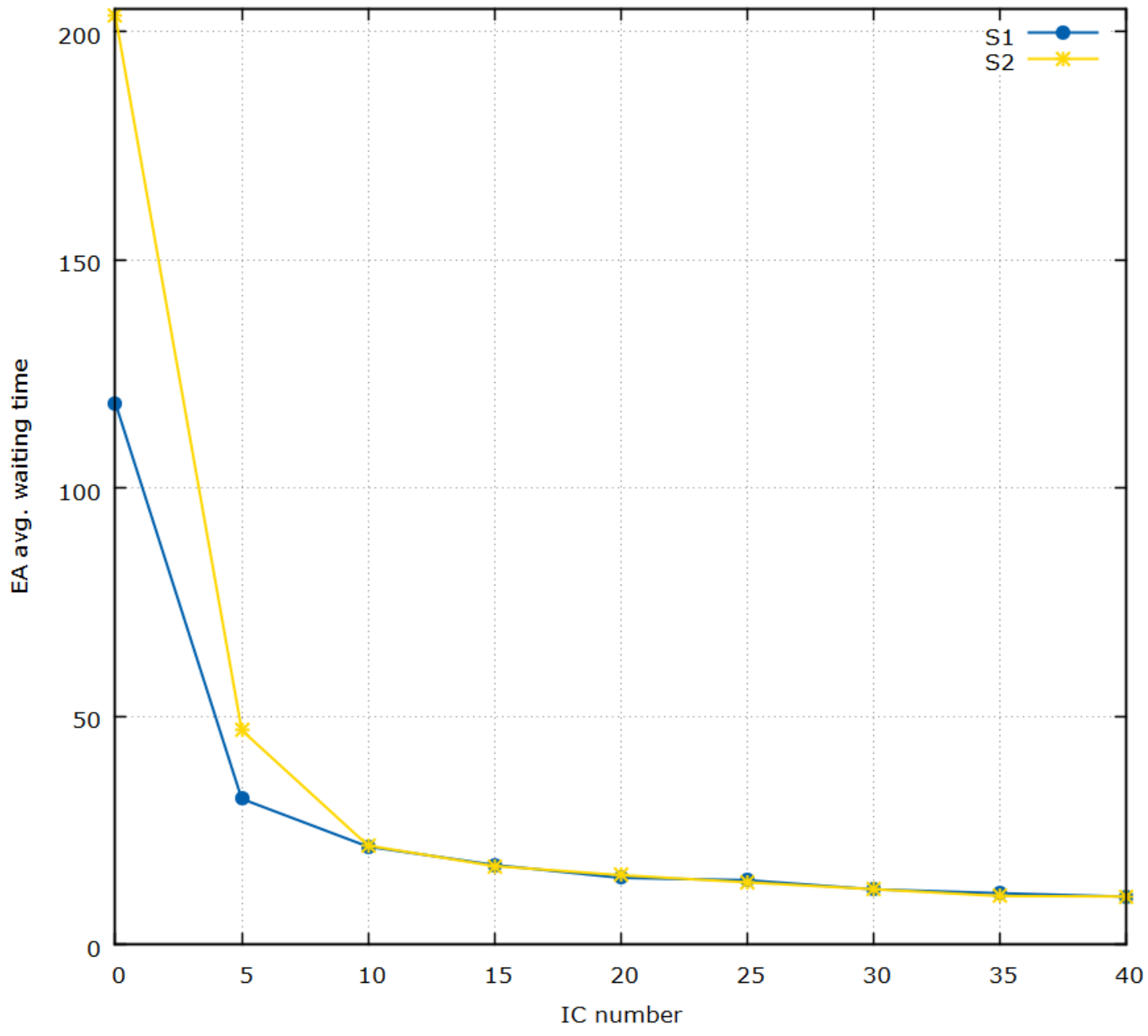

Figure 5.7: Average waiting times with S1 and S2 and a number of IC's ranging from 0 to 40.

Table 5.6: Results of $S_1$ scenarios

| | S1 | S1 - 5 ICs | S1 -10 ICs | S1 - 15 ICs | S1 - 20 ICs | S1 - 25 ICs | S1 - 30 ICs | S1 - 35 ICs | S1 - 40 ICs |
|---|---|---|---|---|---|---|---|---|---|
| FP number | 299 | 320 | 330 | 335 | 357 | 342 | 311 | 344 | 340 |
| FN number | 56 | 63 | 51 | 46 | 74 | 68 | 52 | 48 | 58 |
| TP number | 195 | 212 | 227 | 228 | 221 | 224 | 232 | 238 | 230 |
| TN number | 147313 | 171175 | 165171 | 166988 | 170681 | 167515 | 162385 | 163097 | 166825 |
| Avg. FP/tick | 0.0299 | 0.032 | 0.033 | 0.0335 | 0.0357 | 0.0342 | 0.0311 | 0.0344 | 0.034 |
| Avg. FN/tick | 0.0056 | 0.0063 | 0.0051 | 0.0046 | 0.0074 | 0.0068 | 0.0052 | 0.0048 | 0.0058 |
| FP ratio | 0.6052 | 0.6015 | 0.5924 | 0.595 | 0.6176 | 0.6042 | 0.5727 | 0.591 | 0.5964 |
| FN ratio | 0.000379999 | 0.000367909 | 0.000308676 | 0.000275393 | 0.000433369 | 0.000405769 | 0.000320124 | 0.000294217 | 0.000347549 |
| Sensivity | 77.68 | 77.09 | 81.65 | 83.21 | 74.91 | 76.71 | 81.69 | 83.21 | 79.86 |
| Specificity | 99.79 | 99.81 | 99.8 | 99.79 | 99.79 | 99.79 | 99.8 | 99.78 | 99.79 |
| CSC (ambulances) | 33720 | 25928 | 25230 | 25062 | 23951 | 24333 | 23541 | 24619 | 24102 |
| CSC (volunteers) | 0 | 0 | 11351 | 9252 | 7874 | 7414 | 6023 | 5963 | 5400 |
| CWT | 58617 | 16981 | 11943 | 9815 | 8451 | 7980 | 6565 | 6545 | 5969 |
| Reqs. Handled | 494 | 532 | 557 | 563 | 578 | 566 | 543 | 582 | 570 |
| IC verifications | 0 | 472 | 540 | 555 | 576 | 566 | 540 | 582 | 568 |
| MA verifications | 296 | 33 | 8 | 6 | 0 | 0 | 1 | 0 | 0 |
| MA interventions | 193 | 211 | 225 | 227 | 220 | 222 | 231 | 235 | 230 |
| Avg. MA cost | 68.25910931 | 48.73684211 | 45.2962298 | 44.51509769 | 41.43771626 | 42.99116608 | 43.35359116 | 42.30068729 | 42.28421053 |
| Avg. WT | 118.6578947 | 31.91917293 | 21.44165171 | 17.43339254 | 14.62110727 | 14.09893993 | 12.09023941 | 11.24570447 | 10.47192982 |

Table 5.7: Results of $S_2$ scenarios

| | S2 | S2 - 5 ICs | S2 - 10 ICs | S2 - 15 ICs | S2 - 20 ICs | S2 - 25 ICs | S2 - 30 ICs | S2 - 35 ICs | S2 - 40 ICs |
|---|---|---|---|---|---|---|---|---|---|
| FP number | 418 | 603 | 631 | 629 | 663 | 654 | 661 | 651 | 667 |
| FN number | 20 | 24 | 29 | 29 | 23 | 28 | 31 | 30 | 30 |
| TP number | 178 | 235 | 245 | 257 | 255 | 239 | 256 | 255 | 256 |
| TN number | 103699 | 145952 | 156918 | 155168 | 157907 | 164355 | 160108 | 161965 | 162078 |
| Avg. FP/tick | 0.0418 | 0.0604 | 0.0631 | 0.0629 | 0.0663 | 0.0654 | 0.0661 | 0.0651 | 0.0667 |
| Avg. FN/tick | 0.002 | 0.0024 | 0.0029 | 0.0029 | 0.0023 | 0.0028 | 0.0031 | 0.003 | 0.003 |
| FP ratio | 0.7013 | 0.7195 | 0.7203 | 0.7099 | 0.7222 | 0.7323 | 0.7208 | 0.7185 | 0.7226 |
| FN ratio | 0.000192829 | 0.000164411 | 0.000184776 | 0.000186859 | 0.000145634 | 0.000170334 | 0.000193582 | 0.000185191 | 0.000185062 |
| Sensivity | 89.89 | 90.73 | 89.41 | 89.86 | 91.72 | 89.51 | 89.19 | 89.47 | 89.51 |
| Specificity | 99.59 | 99.58 | 99.59 | 99.59 | 99.58 | 99.6 | 99.58 | 99.59 | 99.59 |
| CSC (ambulances) | 38819 | 30249 | 27141 | 26829 | 26054 | 25832 | 26026 | 25296 | 25116 |
| CSC (volunteers) | 0 | 33426 | 18114 | 14344 | 13048 | 11269 | 10199 | 8723 | 8785 |
| CWT | 121316 | 39364 | 18986 | 15229 | 13965 | 12160 | 11113 | 9628 | 9707 |
| Reqs. Handled | 596 | 838 | 876 | 886 | 918 | 893 | 917 | 906 | 923 |
| IC verifications | 0 | 745 | 824 | 877 | 913 | 887 | 913 | 904 | 922 |
| MA verifications | 420 | 67 | 37 | 5 | 2 | 3 | 1 | 1 | 0 |
| MA interventions | 161 | 219 | 228 | 235 | 237 | 226 | 239 | 232 | 232 |
| Avg. MA cost | 65.13255034 | 36.09665871 | 30.98287671 | 30.28103837 | 28.38126362 | 28.92721165 | 28.38167939 | 27.9205298 | 27.21126761 |
| Avg. WT | 203.5503356 | 46.97374702 | 21.67351598 | 17.18848758 | 15.2124183 | 13.61702128 | 12.11886587 | 10.62693157 | 10.51679307 |

# Conclusions

Evolution is a process that has constantly accompanied the humanity. The same should apply for the traditional organizations. Evolving the traditional organizations while maintaining the identity of the intended services represents a great challenge for the mankind. With an ever increasing number of population and with scarcer natural resources a more intelligent way of management and organization is a must. The time to explore smarter and more efficient ways of organization is now.

In this thesis we presented two simulation models and preliminary results using the FSO concepts as a potential solution to the given problems. It was indicated that FSO provides us with an example of the social-energy-aware solutions we referred to above. The provided simulation models have shown that FSO's dynamic hierarchical organization optimally orchestrates all participating entities by the use of exceptions mechanism, thus overcoming the stiffness of the traditional organizations. By the use of this interconnecting structure of roles of various natures a significantly improved agility is reached. The inclusion of intelligent mobile agents offers enormous collaboration opportunities and resource sharing, thus leading to systems able to tap into the great wells of social energy of our societies.

In the first simulation model we showed that by the use of FSO properties the individuals may benefit by receiving more qualitative healthcare services. This is due to the increase of accuracy of service provider identification. At the same time the service receiving times improved due to the inter-connectivity of organizations, allowing for fast flow of information. By providing a service that involves entities of various organizations a better usage of resources is archived, thus exploiting more of the social energy. The diversity of FSO applications is shown by the number of examples where such approach might have an application. In the second simulation model we showed that by the usage of FSO mechanism fall detection systems can be improved. Depending on the specificity and sensitivity of devices as well as on the willingness and availability of human beings, we showed how it is possible to improve the social costs, make better use of the social resources, and reduce the average time to respond to identified falls.
In the next section we describe the future work.

# Future work

The subjects where FSO approach can get an application are numerous. However, with a limited amount of time, choices for the subjects to be handled had to be made. As the work progressed, several interesting ideas about various simulation models came up. Some of them we did implement, but for the others a substantial amount of effort and time was required thus they were left out.

An improvement to the current simulation models could be the addition of more detailed characteristics to the agents, which would help mimicking the real life scenarios in a more accurate way. By having more information about the underlying agents the community representatives could be improved to provide with semantic analysis when dealing with given situations. Further the emergence of novel and possibly unexpected service modes, where two or more primary agents (elderly people) fulfill each others requirements thus creating a self-serve coupling could be explored. This would allow for analyzing the emergence of social self-service mechanisms and their social impact.

As a future work the existing models could be complemented with protocols, algorithms and components that allow for a practical implementation of the Service-oriented Communities and Fractal Social Organizations. The idea is switching from the current abstract models, to concrete software artifacts. A possible solution could be the implementation of a middleware based on Web Services to manage intra- and inter-SoC cooperation.

The topic is very broad and future research can take many directions.

# Appendix A

# Exemplary FSO structural organization

FSO-structure.txt

```
Emergency response members                                   L2
    firefighter's_coordinator 37                             L1
    people's_coordinator 36                                  L1
    localhospital 8                                          L1
    localhospital 7                                          L1
People's coordinator members                                 L1
    [taxi 160 taxidriver 154]                                L0
    [taxi 164 taxidriver 157]                                L0
    [taxi 165 taxidriver 152]                                L0
    [taxi 170 taxidriver 151]                                L0
    [taxi 171 taxidriver 159]                                L0
    [taxi 172 taxidriver 160]                                L0
    [taxi 173 taxidriver 158]                                L0
    [taxi 159 taxidriver 155]                                L0
    [individual 144]                                         L0
    [individual 139 car 151]                                 L0
    [individual 140]                                         L0
    [individual 142]                                         L0
    [individual 143 car 149]                                 L0
    [individual 138 car 148]                                 L0
    [individual 103 car 199]                                 L0
    [individual 131 car 101]                                 L0
    [individual 161 car 131]                                 L0
    [individual 120 car 166]                                 L0
    [individual 147 car 150]                                 L0
    [individual 146]                                         L0
    [house 116 fire-detector 60]                             L0
    [house 133 fire-detector 56]                             L0
    [house 96 fire-detector 78]                              L0
    [house 104 fire-detector 44]                             L0
Firefighters's coordinator members                           L1
    [ftruck 166 firefighter 172 firefighter 169]             L0
    [ftruck 167 firefighter 171]                             L0
    [ftruck 168 firefighter 170 firefighter 173 ]            L0
localhospital 8 members                                      L1
    medicalappliance 27                                      L0
    medicalappliance 25                                      L0
    medicalappliance 31                                      L0
    medicalappliance 33                                      L0
    medicalappliance 28                                      L0
    medicalappliance 26                                      L0
```

| | |
|---|---|
| doctor 20 | L0 |
| doctor 24 | L0 |
| transportvan 16 | L0 |
| transportvan 13 | L0 |
| transportvan 12 | L0 |
| transportvan 14 | L0 |
| ambulance 10 | L0 |
| ambulance 9 | L0 |
| ambulance 11 | L0 |
| localhospital 7 members | L1 |
| medicalappliance 34 | L0 |
| medicalappliance 30 | L0 |
| medicalappliance 29 | L0 |
| doctor 22 | L0 |
| doctor 23 | L0 |
| transportvan 17 | L0 |
| transportvan 19 | L0 |